\documentclass[preprint2]{aastex}

\input epsf

\newcommand{\Msun}{M_{\odot}}
\newcommand{\Msunh}{h^{-1}M_{\odot}}
\newcommand{\etal}{{\it et al} \ }

\newcommand{\gsim}{ \lower .75ex \hbox{$\sim$} \llap{\raise .27ex \hbox{$>$}} }
\newcommand{\lsim}{ \lower .75ex \hbox{$\sim$} \llap{\raise .27ex \hbox{$<$}} }

\newcommand{\Mpch}{\, h^{-1}{\rm Mpc}}    
\newcommand{\kpch}{\, h^{-1}{\rm kpc}}    
 \newcommand{\kpc}{\,{\rm kpc}}    
 \newcommand{\Mpc}{\,{\rm Mpc}}    
\newcommand{\kms}{\, {{\rm km}}\,{{\rm s}}^{-1} }
 \newcommand{\hub}{ \, {{\rm Km}}\,{{\rm s}}^{-1}\,{{\rm Mpc}}^{-1} }

\newcommand{\be}{\begin{equation}}
\newcommand{\ee}{\end{equation}}

\slugcomment{Submitted to ApJ, September 30, 1999}

\shorttitle{Dark Matter Halos at Ultra-High Resolution}
\shortauthors{Ghigna et al.}

\begin{document}

\title{Density profiles and substructure of dark matter halos: \\
converging results at ultra-high numerical resolution.}

  \author{S.~Ghigna\altaffilmark{1},
  B.~Moore\altaffilmark{1}, F.~Governato\altaffilmark{2},
  G.~Lake\altaffilmark{3}, T.~Quinn\altaffilmark{3},
  J.~Stadel\altaffilmark{3}}

\altaffiltext{1}{Physics Department, University of Durham, Durham City, UK}

\altaffiltext{2}{Osservatorio di Brera-Merate, Milano, Italy}

\altaffiltext{3}{Astronomy Department, University of Washington, Seattle, USA}

\begin{abstract} 

Can dissipationless N-body simulations be used to reliably determine the
structural and substructure properties of dark matter halos? A large
simulation of a galaxy cluster in a cold dark matter universe is used to 
increase the force and mass 
resolution
of current ``high resolution simulations'' by almost an order of magnitude 
to examine
the convergence of the important physical quantities.  
The cluster contains $\sim 5$ million 
particles within the final virial radius, $R_{vir}\simeq 2 \Mpc$ (with 
$H_0=50\hub$), and is simulated using 
a force resolution of $1.0\kpc$ ($\equiv 0.05\%$ of $R_{vir}$); the final
virial mass is $4.3\,10^{14}\Msun$, equivalent to a circular velocity 
$v_{circ}\equiv (GM/R)^{1/2}\simeq 1000 \kms$ at the virial radius.
 The central
density profile has a logarithmic slope of -1.5, identical to 
lower resolution studies of the same halo, indicating that the profiles 
measured from simulations of this resolution have converged to the 
``physical'' limit down to scales of a few kpc ($\sim 0.2\%$ of $R_{vir}$).
Also the abundance and properties of substructure are consistent with 
those derived from lower resolution runs;
from small to large galaxy scales 
($v_{circ}>100\kms$, $m>10^{11}\Msun$), 
the circular velocity function and the mass function
of substructures can be approximated by power laws with slopes $\sim -4$ and
$\sim -2$ respectively. 
At the current resolution, 
overmerging --- a numerical effect that leads to structureless virialized halos
in low-resolution $N$-body simulations ---
seems to be globally unimportant 
for substructure halos with circular velocities $v_{circ} > 100\kms$ 
($\sim 10$\% of the cluster's $v_{circ}$).
We can identify subhalos
orbiting in the very central region of the cluster ($R\lsim 100\kpc$)
and we can trace most of the cluster progenitors from high redshift to the
present.
The object at the cluster center 
(the dark matter analog of a cD galaxy) is assembled between $z=3$ and $z=1$ 
from the merging of a dozen halos with $v_{circ}\gsim 300\kms$.
Tidal stripping and halo-halo collisions 
decrease the mean circular velocity of the substructure halos by $\approx 20\%$
over a 5 billion year period. 
We use the sample of 2000
substructure halos to explore the possibility of biases using galactic 
tracers in clusters: the velocity dispersions of the halos 
globally agree with the dark matter within $\lsim 10\%$, but 
the halos are spatially anti-biased, and in the very central region of 
the cluster ($R/R_{vir}< 0.3$), they show positive velocity bias ($b_v\equiv 
\sigma_{v3D,halos}/\sigma_{v3D,DM} \simeq 1.2$--$1.3$); however, this effect
appears to depend on numerical resolution.

\end{abstract}


\section{Introduction}

In hierachical cosmological scenarios galaxies and clusters form in virialized
dark matter dominated halos that are assembled via merging and accretion of
smaller structures (White \& Rees 1978, Davis \etal 1985; for a recent 
analysis, e.g. Tormen 1997, 1998). Until recently, to what extent
the {\sl subhalos} survive within the potential well of the larger system has
been largely uncertain, because cosmological $N$-body simulations 
were not able to resolve more than a handful of substructure halos
(e.g. Carlberg 1994, Summers, Davis \& Evrard 1995, Frenk \etal 1996).
Infalling
subhalos are heated by tidal forces generated by the global potential
and by mutual encounters and rapidly lose a large fraction of their masses; 
this is a physical effect but is greatly enhanced by limited numerical resolution.  
The finite resolution sets an upper limit to the
potential depth of halos - large force softening or low numbers of particles 
per
halo conspire to produce soft, diffuse substructure 
halos that are easily disrupted by tidal forces (Moore,
Katz \& Lake 1996) and lead to structureless virialised halos. 
This is the classic {\sl overmerging} problem (White \etal 1987).

Gas physics is not a solution; it is necessary to accurately reproduce
the dynamics of the (dominant) dark 
matter component. It is now clear 
that mass and force resolution of the simulations are the key
parameters for overcoming the overmerging problem (Moore, Katz \& Lake 1996,
Brainerd, Goldberg \& Villumsen 1998, Moore \etal 1998, Ghigna \etal 
1998, Tormen, Diaferio \& Syers 1998, Klypin, Gottl\"ober, Kravtsov \& 
Kokhlov 1999a, Okamoto \& Habe 1999).
Increased resolution leads to 
substructure
halos ({\sl subhalos} hereafter) with higher central densities, 
enabling them to survive. 
Halos extracted from large cosmological simulations 
and re-simulated (see next section) with $\sim
10^6$ particles and force resolution $\lsim 0.01\%$ of the virial 
radius yield a wealth of substructure allowing a comparison between 
the mass and light distributions within clusters and galaxies (Ghigna \etal
1998, hereafter G98; Okamoto \& Habe 1999, Moore \etal 1999a). 
Tormen, Diaferio \& Syers (1998) have addressed the same issue using a 
sample of clusters simulated at lower resolution. 
High-resolution simulations of 
moderately large cosmological volumes 
that retain significant amounts of substructure within virialized halos 
have also become recently feasible (Klypin \etal 1999a, hereafter KGKK; 
Klypin, Kravtsov,
Valenzuela \& Prada 1999b; Col\'{\i}n, Klypin, Kravtsov \& Khokhlov 1999;
Col\'{\i}n, Klypin \& Kravtsov 1999; see also Kauffman \etal 1999a, 1999b
and Diaferio \etal 1999). These latter approaches have the advantage of
providing relatively large samples of dark matter halos representing clusters, 
groups or galaxies, but cannot detect systematic biases introduced by the
limited resolution.
 
The central density profile of halos is also affected by numerical resolution.
In order to compare the predictions of hierarchical models with observational
data on the mass, light and X-ray profiles of clusters
(e.g. Carlberg \etal 1996, Carlberg, Yee \& Ellingson 1997, Smail \etal 1997,
AbdelSalam, Saha 
\& Williams 1998, Adami, Mazure, Katgert \& Biviano 1998, Allen 1998, 
Markevitch, Vikhlin, Forman \& Sarazin 1999),
galaxy rotation curves (e.g. Moore 1994, Flores \& Primack 1994, 
Moore \etal 1999b and references therein), giant arc properties in 
gravitationally lensing clusters
(e.g. Kneib \etal 1996; also, 
Williams, Navarro \& Bartelmann 1999, Flores, Maller \& Primack 2000,
Meneghetti \etal 2000) or
constraints on processes directly related to the nature of the
dark matter, such as particle-particle
annihilation rates (Calcaneo-Roldan \& Moore, in preparation), it is 
important to resolve the central 
structure
of dark matter halos.  
Using $N$-body simulations with $10^4\--10^5$ particles per halo 
for the CDM models and its popular variants,
Navarro, Frenk \& White (1996, 1997) found that
the profiles of isolated relaxed dark matter halos 
can be well described by a ``universal'' profile (NFW profile)
from galactic to cluster scales; these results have been confirmed by other 
authors using simulations of comparable resolution --- e.g Cole \& Lacey 1996, 
Tormen, Bouchet \& White 1996.
However, improving the numerical resolution (Moore \etal 1998) 
leads to profiles with 
central cusps significantly steeper than that of an NFW profile 
($\rho(r)\propto r^{-1}$ for the latter); halos simulated with 
$\sim 10^6$ particles have profiles fit by the functional form 
$ \rho(r) \propto [(r/r_s)^{1.5}(1+(r/r_s)^{1.5})]^{-1}$
(Moore \etal 1999b),
which has a cusp $\rho(r) \propto r^{-1.5}$ 
as $r\rightarrow 0$. 
(In a recent analysis, Jing \& Suto, 2000, find similar results
for galaxies and groups, but
shallower central profiles for a sample of clusters
simulated in a $\Lambda$CDM cosmology; see our comments in \S~4).

In this paper, we examine how much resolution per halo is required to
make  numerical effects negligible for various physical quantities
and obtain robust results on the halo density profiles and
the space, mass, velocity distribution of substructures. 
We perform one large (and expensive) simulation of a dark matter halo,
taking the simulation originally analysed by Ghigna
\etal (1998; hereafter G98) 
and increasing the force and mass resolution by almost an order of 
magnitude.

The plan of the paper is as follows: In \S 2, we describe the N-body
simulations and, in \S~3, the method used to identify the substructure halos. 
In \S~4, we consider the issue of the typical density profile of isolated and 
substructure halos. Section~5 is 
devoted to
the statisitcal properties of the substructure, the effects of resolution,
evolution and environment. We study the distribution of their internal
velocities and masses, their spatial distribution and whether they trace the
mass, and also the issue of velocity bias. In \S~7, we 
examine the 
relation between the cluster progenitor halos at high redshift and the substructure halos 
at the present epoch, and the formation of the
central ``cD'' halo. We summarise the results and conclude in \S~8.
\label{s:Intro}

\section{The $N$--body simulations}

Our aim is to achieve very high spatial and mass resolution within the dark 
halo of a cluster drawn from a ``fair volume" of a standard CDM universe.  
To achieve
such resolution, we initially perform a simulation of a large volume ($50\Mpch$
per side) of a CDM universe (normalised such that $\sigma_8=0.7$ and with the
shape parameter $\Gamma=0.5$ and $H_0=50\hub$).  The
size of this box is nearly ten times the turnaround radius allowing the tidal
field and hence infall pattern to be modelled correctly.  We choose a cluster
that is virialised by a redshift $z=0$ and use a technique of ``volume
renormalization" to obtain higher resolution within the region of interest (see
G98 and references therein). Our goal is
to simulate the formation of this cluster such that $\sim 10^7$ particles lie
within the turnaround radius at the final time.  (Note that if we simulated the
entire volume at this resolution we would need to use a total of $6\times 
10^8$ particles.) Beyond the turnaround region the mass
resolution is decreased in a series of shells allowing us to reproduce the
external tidal field correctly.  
The particle distribution is evolved using the
high performance parallel treecode PKDGRAV, using periodic boundaries and a
variable timestep scheme based upon the local acceleration (Quinn {\it et al}
1997, Wadsley, Quinn \& Stadel 2000, in preparation). 
To maintain accurate forces when the mass distribution is fairly regular, 
from a redshift z=69 to z=2 we use an opening angle of $0.4$ and we complete 
the simulation using an opening angle of 0.7. We 
expand the cell moments to hexadecapole order. The code uses a spline 
softening length such that the force is completely Newtonian at twice our 
quoted softening lengths.
The time integration method is 
an adaptive implementation of the
standard leapfrog integration scheme. 
Individual particle time steps $\tau$
are chosen to satisfy the criterium 
$\tau \lsim \eta(\epsilon/a)^{1/2}$, where $\epsilon$ is the
force softening length of a particle, 
$a$ the magnitude of the local acceleration
and $\eta$ a constant determined on the basis of
stability and accuracy requirements (for these runs $\eta = 0.2 $).  
In order to
maintain
synchronization of the system, the time steps are quantized
on power-of-two subdivisions of the largest
time step (the ``system'' time step $\tau_s$), i.e. $\tau \equiv \tau_n = \tau_s / 2^n $ (e.g., Hernquist \& Katz 1989). For these runs, 
500--1000 system steps are used, equally spaced in time. 

Here we consider two simulations of the cluster (the same object as in G98)
with largely different resolutions, which we label LORES and HIRES.
In the high resolution region of the box, the
particle mass is $4.3\times10^8\Msunh$ for LORES
and $5.37\times10^7\Msunh$ for HIRES;
the softening is 1 and $0.5\kpch$ respectively. 
At the final epoch, the cluster has a radius $R_{200} = 1.0\Mpch$ 
($R_{200}$ is the
radius within which the average density is 200 times the cosmic density and it
is a measure of the extent of the virialized region) and a mass within such
radius $M_{200}= 2.15\times10^{14}\Msunh$.  With
these parameters, run HIRES has more than 4 million particles within
$R_{200}$, a factor of 8 of improvement with respect to LORES.

\begin{figure}
\centering
\epsfxsize=5truecm\epsffile{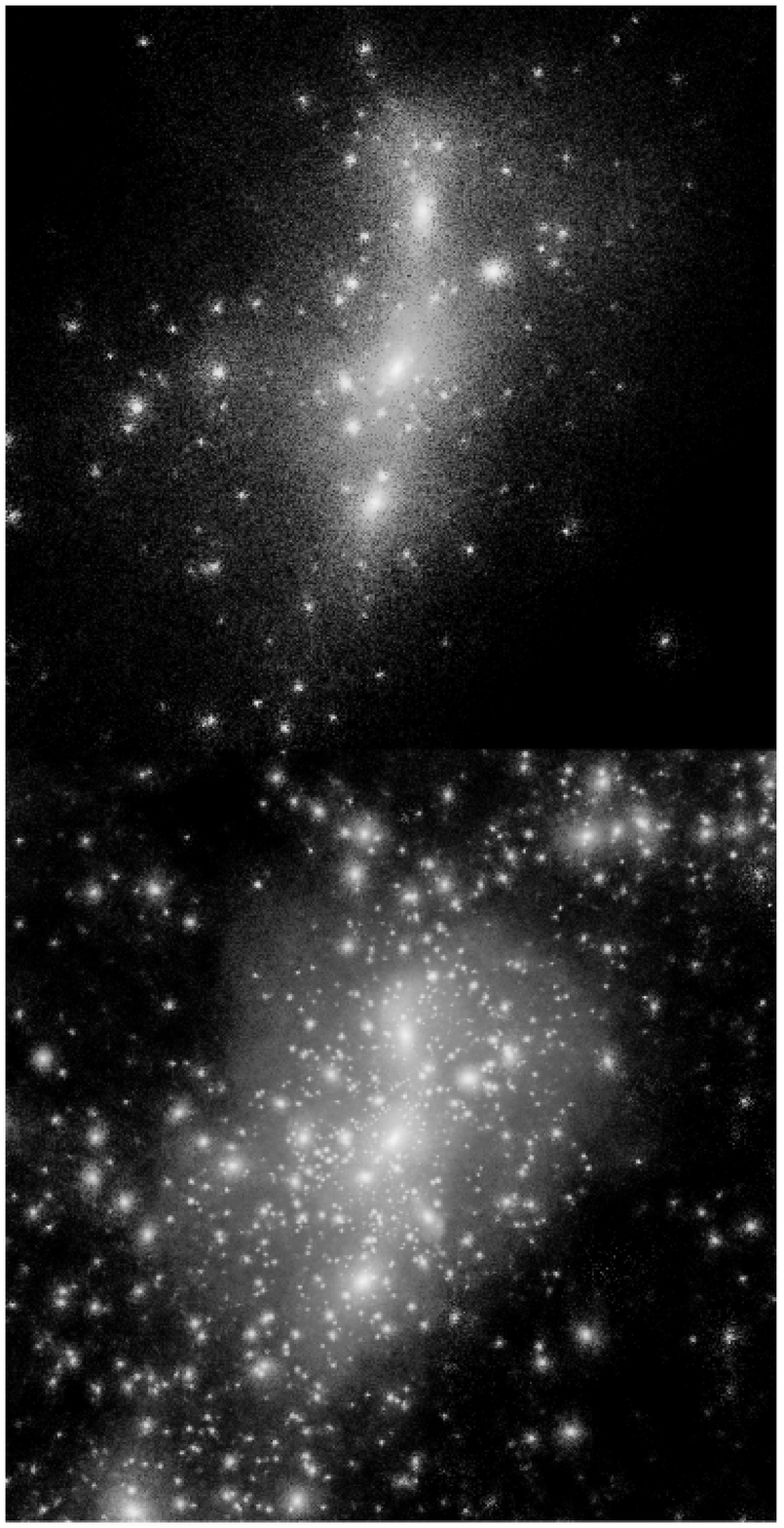}
\caption{Maps of the cluster's density as seen in 
run LORES (upper panel) and HIRES, which has 8 times better
mass resolution. The change in the appearance between the two runs
is mainly due to HIRES's ability of resolving further down the  
substructure mass function.
} 
\label{f:2maps}
\end{figure}

The mass growth curve of this cluster is shown in Figure~3 of 
Ghigna \etal (1998). 
The cluster does not undergo major mergers since $z=0.5$;
the virial mass increases by $\lsim 30$\% since that epoch, while the mass
within the physical volume of the $z=0$ cluster (i.e. the mass 
within $R<R_{200}|_{z=0}$) grows by only $\sim 10$\%. 
The early formation and ``quiet'' recent history of the cluster are 
convenient for this study, 
since they allows us to single out the effects of tides 
from major accretion or merger events. The results presented here are 
very likely typical of observed well relaxed clusters, but cannot be easily
generalized to objects with more turbulent recent histories. 

\label{s:nbody}

\section{Halo substructure and its identification}

The wealth of substructure that exists within the cluster is exemplified by
Figure~\ref{f:2maps}.  The panels show maps of the projected density
distribution in a box of side $2R_{200}$ for LORES (upper) and HIRES (lower
panel).  Each particle is plotted using a grey scale according to the logarithm
of the local density (defined using an SPH smoothing kernel over 64 neighbouring
particles using the code {\sl SMOOTH} of Stadel \& Quinn 1997, http ref:
http://www--hpcc.astro.washington.edu/tools).  Only regions with density
contrast $\delta>30$ are shown.  The change of resolution from LORES to HIRES
has a dramatic effect on the abundance of substructure: lower mass
objects are resolved throughout the cluster and features appear in the 
very central region which was previously smooth.

The density maps are obtained at redshift $z=0.5$, at which epoch more than 
80\%
of the mass of the cluster is in place, yet it is not fully relaxed.  In
Figure~\ref{f:2maps}, we can see three dominant substructure clumps (center,
north, south-west).  Similar features are
observed in the light and X-ray distribution of observed high-redshift clusters
(e.g. Abell 521; Maurogordato \etal 2000).  By $z=0$, the second largest 
clump merges with the central object, the dark counterpart of a cD galaxy.
At $z=0.5$, the cluster's radius and masses are $0.6\Mpch$ (physical units) and
$1.7\times10^{14}\Msunh$.

Identifying the substructure halos is a critical step 
(G98, KGKK). Here we use the group finding algorithm SKID,
developed by Stadel \etal (for a full description of the code see
http://www--hpcc.astro.washington.edu/tools). SKID is a ``Lagrangian'' version
of DENMAX (Gelb \& Bertschinger 1994), 
the densities being evaluated at each particle position using
an SPH smoothing kernel rather than at the nodes of a grid.  The particles are
moved along the density gradients until each oscillates around a point.  They
are then linked using a friends-of-friends algorithm (FOF; Davis \etal 1985) 
and the groups (halos) found are checked iteratively for self-boundness.

In G98 we have verified the efficiency of SKID and tested the robustness of
its estimates of the halo structural parameters (see below) against alternative
methods.  The ability of SKID to single out particles bound to substructure from
the diffuse particles bound solely to the entire cluster (smooth background
particles) depends crucially on the linking length, $l_{FOF}$, adopted for the
FOF stage. If $l_{FOF}$ is too small a subhalo will be fragmented into smaller
units whereas if it is too large then halos will be merged together.  A value
$l_{FOF} \simeq 1/3 r_{peak}$, where $r_{peak}$ is the distance from the halo's
center at which the circular velocity profile has its peak $v_{peak}$ (see,
e.g., G98, section~3.3), is usually satisfactory. In our simulations, the
subhalos have a large range of sizes and there is not a value for
$l_{FOF}$  that
is ideal for all of them at a time; moreover there are many instances of 
substructure halos that contain their own substructure -- ``halos within halos
within halos''.  
To account for the complete hierarchy of substructure present in the current
simulation we run SKID with three values of $l_{FOF}= 1.5$, 5 and
$10\,l_{soft}$, and combine the outputs avoiding duplication. 
(SKID run with $l=10l_{soft}$ provides the first level of the
``hierarchy'' of halos; we then add the second level obtained with 
$l=5l_{soft}$ but discarding those halos that are separated from a
first-level halo by a distance less than the latter's $r_{peak}$; and so
forth for the third level. 
The completeness of the final
halo catalog in several regions of the cluster was verified against the density
concentrations visible in the 3D density map). However, it turns out that
for the statistics of intermediate-to-massive halos (e.g. those with circular 
velocities larger than $\sim 100\kms$) a run of SKID with $l_{FOF}=10l_{soft}$
is adequate (and fast).

\begin{figure}
\centering
\epsfxsize=6.0truecm
\epsffile{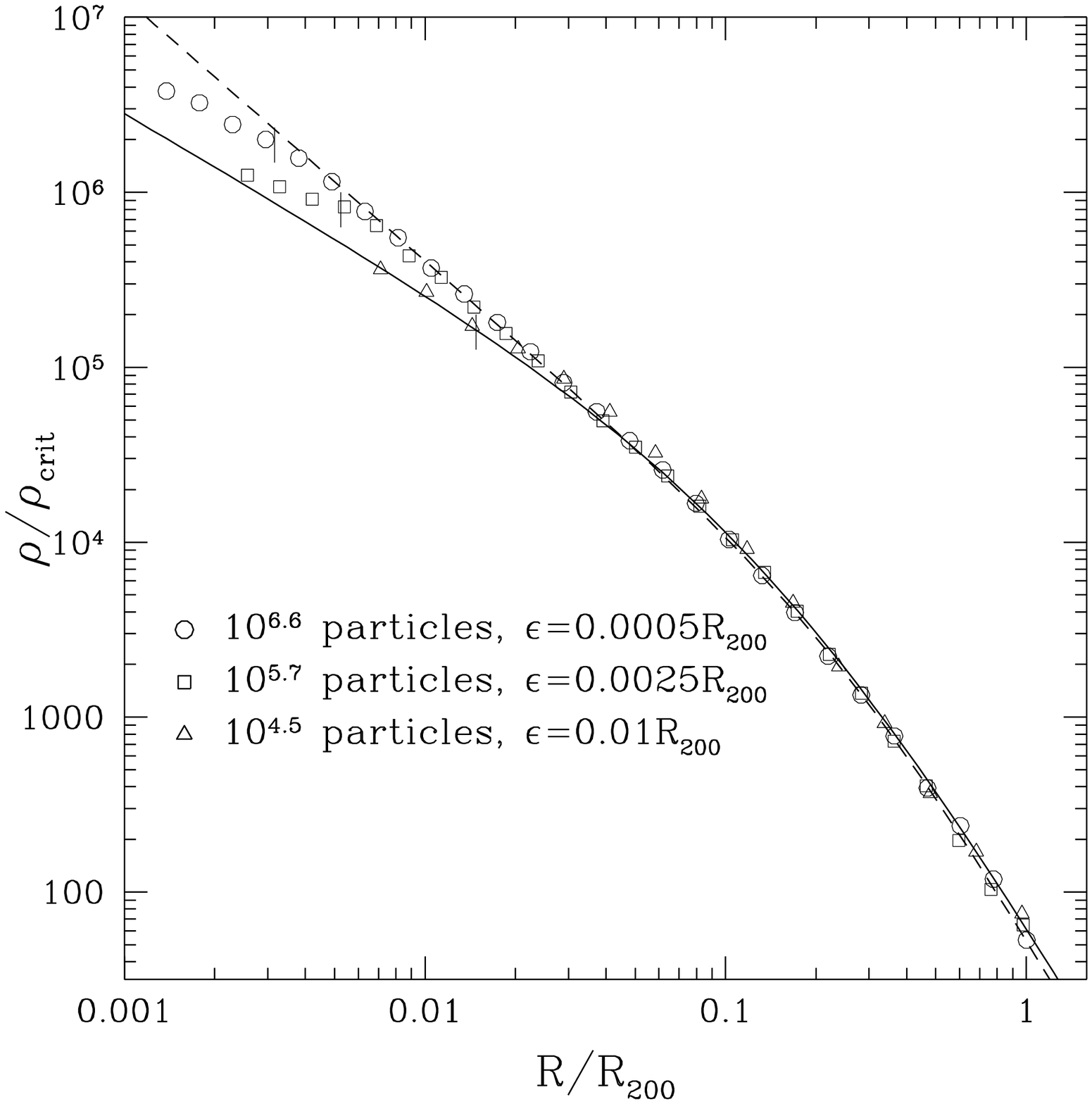}
\caption{The density profile of the cluster measured in three runs with
increasing resolutions (from triangles to squares to circles).  In the best run,
the cluster contains over 4 million particles and the force resolution is
$0.05$\% of the cluster's virial radius. The curves are an NFW profile (lower
curve) and a fit with the profile of Moore \etal (1999a), which rises more
steeply ($\propto r^{-1.5}$) at the center than the NFW profile ($\propto
r^{-1}$).  With increasing resolution, the cluster's profile continues
to approach M99a's curve
$i.e.$ this appears to be the asymptotic profile in the limit of
infinite
 resolution. The
vertical bars mark the radii at which the measured profiles are no longer
affected by finite numerical resolution.} 
\label{f:profiles}
\end{figure}

We consider halos with a minimum number of member particles of 16 
(corresponding to a mass of $8.6 \cdot 10^8\Msunh$ in HIRES). In general, we 
use halos with more than 16 particles when they are employed as tracers, 
but we adopt a minimum
number of 32 particles when their individual properties are relevant.  The high
resolution region analyzed is roughly the final turn-around radius, about twice
the virial radius.  Within this region, in HIRES, we identify over 2000
substructure halos.  Within $R_{200}$, there are $\sim$ 1200 halos at the final
epoch and $\sim 1000$ at $z=0.5$.  The innermost halo is at $45\kpch$ from the
cluster's center although about ten halos lie within a projected distance
of $45\kpch$.

We use the output of SKID to determine the halo structural parameters: radius
$r_{halo}$, mass $M_{halo}$ and ``circular velocity'' $v_{circ}$ (the latter is
the peak value of the rotation curve $v(r) = (M(r)/r)^{1/2}$, where $M(r)$ 
is the halo mass
within a distance $r$ from its center).  
In \S~4.1 of G98, we 
discussed the effects of numerical resolution on the completeness of subhalo
samples.  We modelled the substructure halos as isothermal spheres embedded in a
larger isothermal potential --- the cluster's halo --- and used the standard formula
for tidal stripping by the cluster's potential to estimate the limiting radii
(tidal stripping is the main cause of subhalo disruption, once the cluster is in place). 
(Using NFW profiles to model the halos would not change our estimates 
significantly; see KGKK for a semi-analytical calculation using
NFW profiles). 
 Applying this simple
model to the present simulation indicates that the halo samples obtained from
LORES and HIRES should not be affected by major incompleteness  for 
substructure halos with $v_{circ}\gsim$~100 and $60\kms$,
respectively, assuming pericenters 
greater than $50\kpch$. We take these values as {\sl
quasi-completeness limits}
of the subhalo samples; they mark the values of $v_{circ}$ below which 
 the samples become apparently incomplete.
These estimates do not take into account halo disruption
at redshifts $z>1$, before the cluster is assembled. 
For instance, following the cluster's progenitors from high redshift to the 
present shows 
that the sample derived from HIRES may be $\sim 20$\% incomplete starting 
from $v_{circ}\sim 100\kms$ (see \S~7).

\label{s:iden}

\begin{figure}
\centering
\epsfxsize=6.0truecm
\epsffile{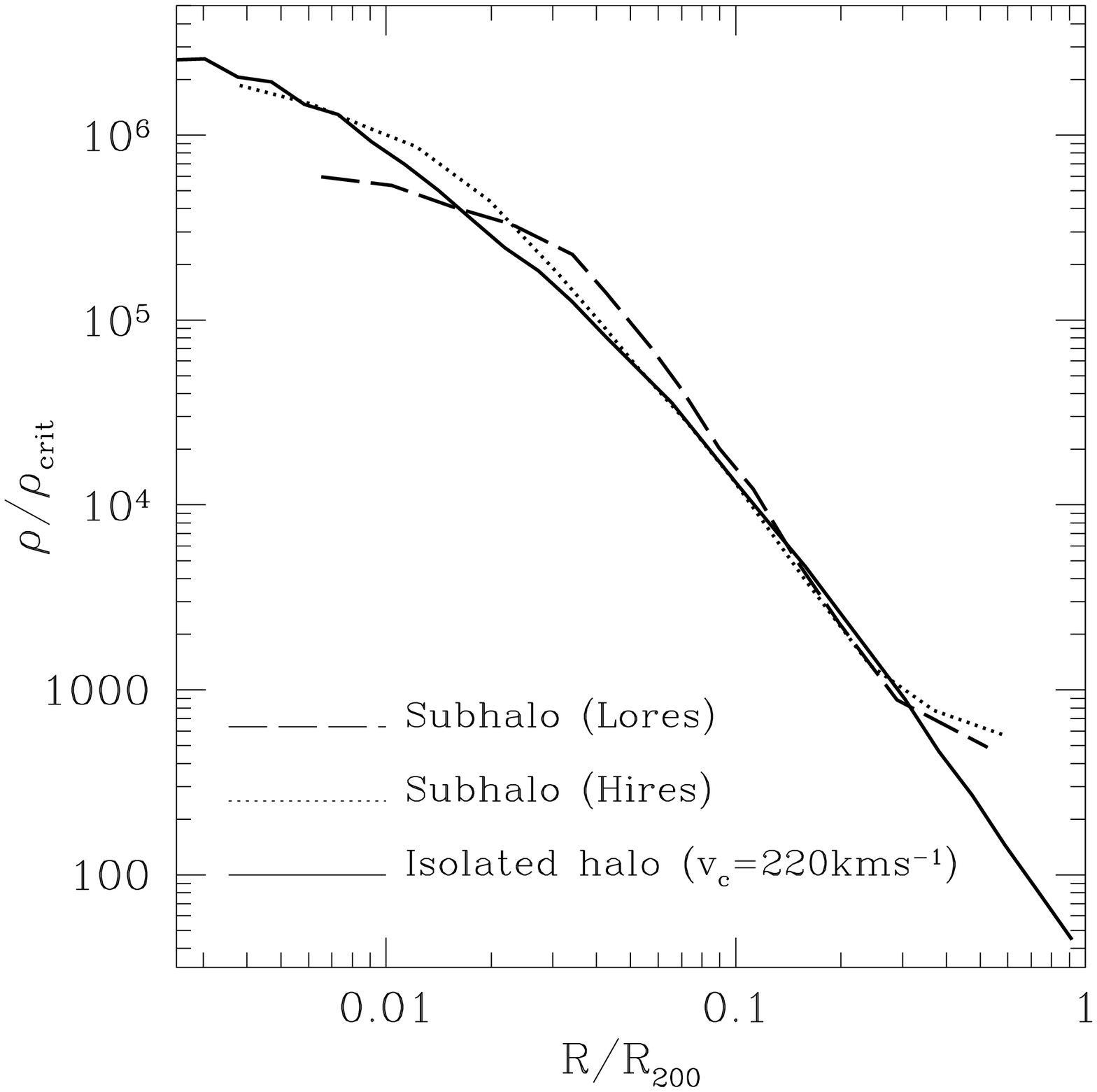}
\caption{The typical density profile of a substructure halo with $v_{circ}\sim
200\kms$ (and clustercentric distance $\sim 0.5\,R_{200}$) in the two runs, 
compared with the profile of an isolated halo (simulated with much
higher resolution). 
Increasing the resolution brings the profiles of substructure halos closer to those
of their isolated counterparts.
}
\label{f:subprofiles}
\end{figure}

\section{The density profile of isolated halos}

One of the most fundamental properties of dark matter halos is their global
density profile. $N$-body simulations have been extensively used to study the
structure of cold dark matter halos, yielding a variety of important results
(e.g. Quinn \etal 1986, Frenk \etal 1988, Dubinsky \& Carlberg 1991, 
Carlberg 1994, Navarro, Frenk \& White 1996, 1997, Cole \& Lacey 1996, 
Tormen, Bouchet \& White 1996, Fukushige \& Makino 1997, Brainerd \etal 1998,
G98, Tormen, Diaferio \& Syer 1998, Huss, Jain \& Steinmetz 1999, 
KGKK, Okamoto \& Habe 1999).  
Systematic studies of halos across a wide range of masses has revealed
a remarkable one parameter scaling (NFW profile; Navarro, Frenk \& White 1996,
1997), with a density profile of the form
$\rho(r)\propto r^{-1}(1+r/r_s)^{-2}$, where $\rho(r)$ is the 
spherically-averaged density at the distance $r$ from the halo's center
and $r_s$ a scale radius.  The halos studied by NFW had
$\approx 10^4$ particles within their virial radii and were resolved to a scale
of a few \% of this region. Moore \etal (1998, 1999b) showed that increasing the
resolution by about an order of magnitude produces steeper central cusps, 
$\rho(r)
\propto r^{-1.5}$. Is the current resolution sufficient to resolve the
central density profile of CDM halos or will increasing the resolution 
make the cusp even steeper?

Figure~\ref{f:profiles} shows the density profile of the cluster, measured in
three runs with different resolutions --- $\sim 10^{4.5}, 10^{5.7}$ and 
$10^{6.6}$  particles
within the final virial radius.  The lower solid curve shows the expected NFW
curve for a halo of this mass in this cosmology. It provides a good fit to our
lowest resolution simulation but underestimates the central density of the
higher resolution runs.  The upper solid curve shows the slightly modified
profile proposed by Moore \etal (1999;
henceforth M99a), $ \rho(r) \propto [(r/r_s)^{1.5}(1+(r/r_s)^{1.5})]^{-1}$,
where $r_s$ is a scale radius. With increasing resolution, 
the cluster's profile appears to converge (asymptotically) to M99a's curve.  
The asymptotic profile is attained in the regions that have many particles 
and beyond several softening lengths; as it is shown by the
vertical bars, density 
profiles measured from these $N$-body simulations can be ``trusted'' only
for scales $\gsim 6$ times the force resolution. 
This statement is likely to apply to simulations run with
different techniques, provided force and mass resolutions are well
balanced, but a detailed study of a large number of halos 
systematically varying the numerical parameters, 
time-stepping criteria etc, would be necessary to establish this
(see also Moore, Katz \& Lake 1996 and Moore \etal 1998). 

With increasing resolution, the profiles of substructure halos 
become closer to those of isolated halos. Figure~3 compares the ``typical''
profile of subhalos with $v_{circ}\sim 200 \kms$ in LORES and HIRES
(obtained for each run by averaging the profiles of three subhalos
with $R\sim 0.5 R_{200}$ and $v_{circ}$ close to 200$\kms$) with the
profile of a similar isolated halo, simulated with 10$^6$ particles
(data taken from Moore \etal 1999). 
It is interesting to note that the typical subhalo profile is
steeper in LORES than in HIRES, perhaps as a result of numerical
relaxation between the hot halo particles and cold subhalo particles.
This confirms that 
caution is required when studying the properties of the profiles
of substructure halos even in high-resolution simulations (G98; 
Avila-Reese \etal 1999, Bullock \etal 1999). 

In a recent paper,
Jing \& Suto (1999) claim that the central density profiles of four
simulated CDM clusters are closer to -1.1 than -1.5 as found here.
 We suspect that this difference may be due to the resolution.  Jing etal
use a Plummer softening with quoted resolution $0.005r_{200}$ (their Figure~2)
and they measure the central slopes using the data between $0.007< r/r_{vir} 
<0.02$. This region may be dominated by dynamics on scales of order the
softening (as seen above, we find that the profiles are reliable at scales
that are 3 times the scale at which the force becomes Newtonian 
-- our spline softening is completely Newtonian at twice the
quoted values whereas a Plummer softening is still ``softer'' than Newtonian
at the resolution quoted by Jing \& Suto).
It should be noted that a steep central
profile (with $\rho(r)\propto \sim -1.4$) 
was also found by Moore \etal 1998 using a
 high resolution simulation
of a different cluster (the ``Coma'' cluster).
At the resolutions achieved here and in Jing etal,
only a handful of systems have been studied and the same cluster has
not been simulated using different codes. We have placed our initial
conditions on our website (www.nbody.net) so that other researchers
can run this cluster and compare with the results presented here and
in Ghigna \etal (1998).

\section{The substructure of dark matter halos}

\begin{figure}
\centering
\epsfxsize=6truecm
\epsffile{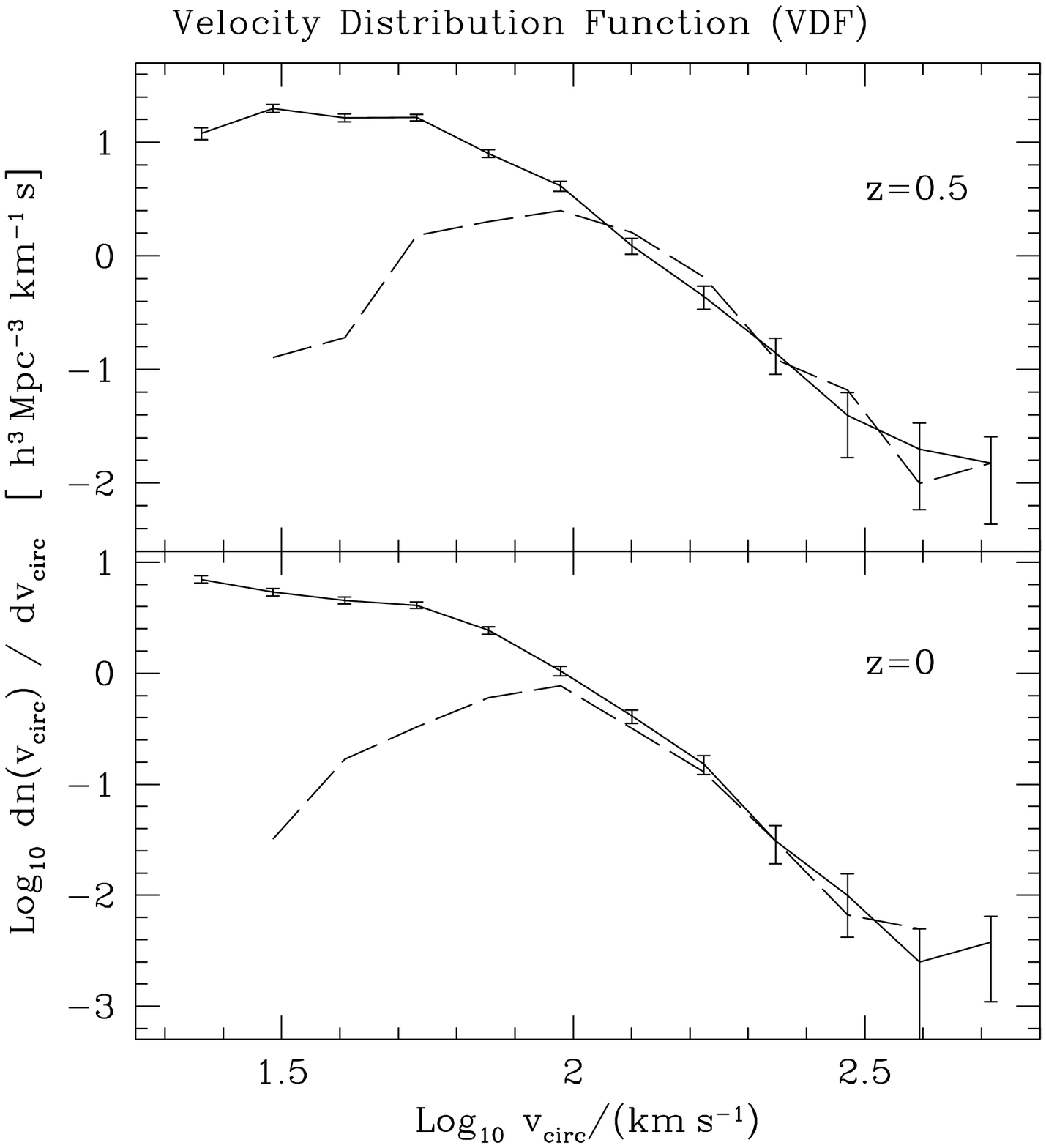}
\caption{Velocity distribution function (VDF) of the substructure halos, at two
redshifts (lower panel: the final epoch, $z=0$; upper panel: the young cluster at
$z=0.5$, when it is $\sim 1$ billion years old). The figure shows the effect of
increasing the resolution by a factor of 8 in mass (from LORES, dashed curve,
to HIRES, solid curve).  The errorbars represent (1-$\sigma$) Poisson errors
on the counts.  The curves agree well where we expect both runs to be
close to completeness (roughly $v_{circ}>100\kms$). The fall off at the low 
velocity end is caused by finite numerical resolution.}
\label{f:dvf2res}
\end{figure}

\subsection{Distribution of circular velocities}

The ``circular'' velocities $v_{circ}$ of the subhalos are the quantities 
that can be most easily compared with observations (although the relation
between 
$v_{circ}$ and the rotation curve or velocity dispersion of the visible
component of a galaxy is not straightforward; e.g. Navarro, Frenk \& White 
1996). 
 Tidal truncation usually changes the mass of a halo by a large amount
(often by $\sim 80$-90\%),
however the peak circular velocity is a fairly stable quantity
and can be regarded, to some extent,
as a ``label'' that can identify a halo for several billion years within the
cluster environment.

\begin{figure}
\centering
\epsfxsize=6truecm\epsffile{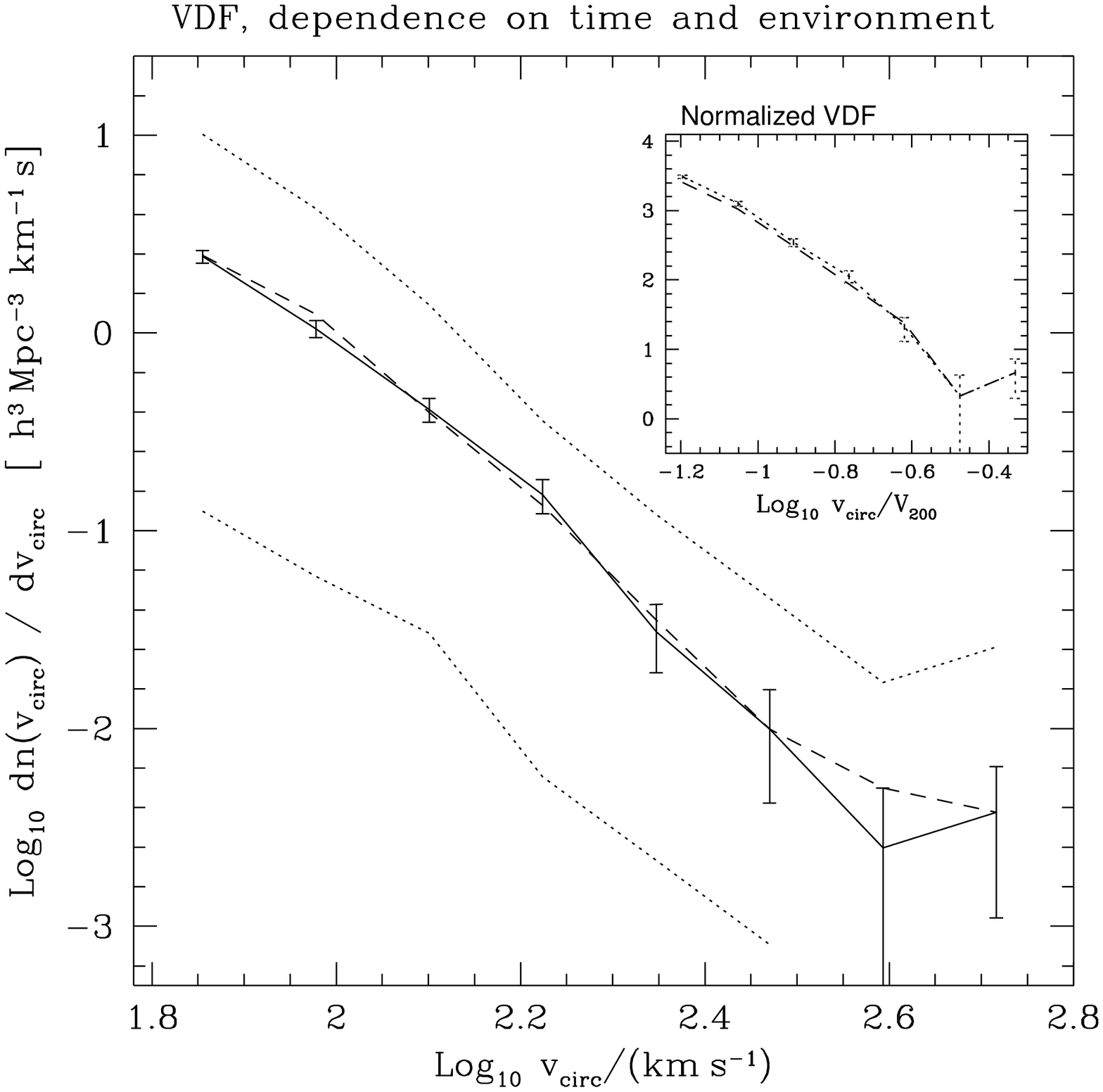}
\caption{The VDF for the cluster's substructure halos at $z=0$ (solid line)
compared with that of the halos at $z=0.5$ contained within the same physical
volume (dashed), using in both cases $R_{200}|_{z=0}$ as limiting distance
(main panel). 
The inset shows the VDFs obtained considering only the halos within 
the cluster's virial radius at $z=0$ (dashed) and $z=0.5$ (dotted) and 
and measuring $v_{circ}$ in units of the cluster's ``circular velocity''
at the respective epoch. In both cases, there are no significant changes 
in the shape and
amplitude of the VDF during the lifetime of the cluster.
The two dotted curves in the main panel show the effect of changing 
environment inside and around the
evolved cluster ($z=0$); they are for inner subhalos ($R/R_{200} < 0.5$; upper
curve) and ``peripheral'' halos ($1 < R/R_{200} < 2.5$; lower). Within and
around the cluster the shape of the VDF is very similar.}
\label{f:dvf_zs}
\end{figure}

Major or total halo disruption is significant for those few massive
halos (with $v_{circ} \gsim 400\kms$) that take part in the formation 
of the cluster and undergo
similar-mass mergers during the early phases of the assemblage, at $z\sim2$
(see \S~7);
once the cluster is in place, 
the dynamical friction timescale is much larger than a Hubble time 
even for halos with mass $\sim 10^{13}\Msun$
(for recent analyses on dynamical friction see van den Bosch \etal 1999 and 
 Colpi, Mayer \& Governato 1999). 
Most of the substructure
halos with small to intermediate masses survive within the cluster.
On average, their central masses (the mass within
$r_{peak}$) do not change dramatically and $v_{circ}$ varies only as the 
cubed root of the mass.
Over 5 billion years, $v_{circ}$ varies by $\sim 20$\% and mainly for halos 
that spend a large fraction of their orbital period in the inner region
of the cluster (G98, KGKK; see also below). 
Furthermore, since mergers are rare in clusters (G98, Okamoto \& Habe 
1999), most halos conserve their identities.
This fact is useful, for example, to compare the $v_{circ}$ 
distribution of a cluster's
substructure halos with that of their progenitors predicted using the
Press-Schechter (PS) approximation.

The differential velocity distribution function (VDF), defined as the number of
halos per unit velocity interval per unit (physical) volume, is shown in
Figures~\ref{f:dvf2res} and \ref{f:dvf_zs}. The first Figure shows the
effect of
changing resolution. It compares the curves obtained for HIRES (solid line)
and LORES (dotted) at the final epoch ($z=0$, upper panel) and for the young
cluster ($z=0.5$, lower panel).  We have used all the halos contained within the
virial radius, $R_{200}$, at the two epochs, $0.975$ and $0.6\Mpch$ respectively.
The fall off at the low-velocity end are caused by incompleteness due to limited
resolution, which dominates HIRES and LORES samples for $v_{circ} < 60\kms$
and $<100\kms$, respectively.  LORES's curve appears to be slightly affected 
by resolution up to $v_{circ}\sim 150\kms$; beyond that $v_{circ}$,  
the results
for the two runs agree well.  
In that range, the VDF at $z=0$ is reasonably well 
approximated by
a power-law with exponent $-4$, motivated by the PS approximation for
field halos in this mass range (e.g. Gross \etal 1998, Tormen 1998, 
Jenkins \etal 1999; Klypin \etal 1999b).

In order to single out evolutionary effects, we compare directly the halo 
distributions for run HIRES  at the two epochs in Figure~\ref{f:dvf_zs}.
In the main panel of the figure, we compare the VDFs measured for the
halos within the {\sl same physical (or proper) volume}, 
as determined by the virial radius of 
the cluster at $z=0$. 
There is virtually no 
difference between the curves for the two redshifts (solid and dashed curves);
the VDF is essentially ``frozen'' in time. 
We find the same result 
for the {\sl normalized} VDFs, that is, the VDFs obtained considering 
only the subhalos within R200 at each respective epoch and  
measuring circular velocities and distances in units of the cluster's 
circular velocity and virial radius, also at the respective epoch; the 
comparison of the two curves is shown in the inset of the figure. 
(For these plots, we only show data for $v_{circ}$
greater than the quasi-completeness limit of HIRES.)

The amplitude of the VDF of the subhalos (those within $0.3$-$0.5 R_{200}$)
decreases a little over time, corresponding to a decrement of
$v_{circ}$ of about 20\%. Tidal stripping removes particles orbiting at the
outskirts of substructure halos but tidal heating of the halo cores, from tidal
shocks near pericenter or from halo-halo encounters may 
also affect their concentrations and accelerate particles to 
the halo boundaries.  
A 20\% decrement of $v_{circ}$ corresponds to a $\sim 50$\%
decrement of the mass.  This effect may be physical but can
be enhanced by limited resolution.  In fact, the LORES simulation
exhibits excessive mass loss 
of the inner halos with medium-to-small masses if compared with HIRES 
(see next section). KGKK have examined the central mass loss 
of a model subhalo of $v_{circ}\sim 200\kms$, orbiting in a rich cluster
with pericenter $\sim 150\kpch$, varying the mass and spatial resolution
(see their \S~2.4).
The halo's 
central mass loss depends on the force softening used, if the latter
exceeds a certain ``optimal'' value ($\sim 1/30$ of the halo's tidal radius at
pericenter, for a Plummer softening; interestingly, the dependance on the 
particle mass is very weak);
however, beyond the ``optimal'' resolution, 
the time evolution of the central mass appears
to converge to a step function which drops by 20-30\% at every passage
at pericenter, yielding a mass loss of $\sim 50$\% over 5 billion years
(the orbital period in their model is 2 Gyrs). 
	In the HIRES simulation,
 the ``optimal'' resolution indicated by KGKK's test 
is met e.g. by 
subhalos of $v_{circ}\sim 100\kms$ at $r_{peri}\sim 70\kpch$ and
subhalos of $v_{circ}\sim 200\kms$ at $r_{peri}\sim 50\kpch$ (we use
the isothermal model approximation described in \S~3 to obtain these 
estimates). The fractions of 
inner subhalos with $v_{circ} = 100$ and $200\kms$ and with estimated 
pericenters less than $70\kpch$ and $50\kpch$ respectively are 
$\sim 
20$\% in both cases; therefore the decrease of the VDF from $z=0.5$ to $z=0$
may be affected by numerical resolution, but, since the corresponding 
mass loss is close to the value expected from KGKK's test, 
it is probably 
a physical effect. 
In any case, it sets an upper limit. 
(For comparison, in LORES, the ``optimal'' resolution for $r_{peri}\sim 
70\kpch$ is met only if  $v_{circ} \gsim 200\kms$).     

The shape of the VDF does not depend significantly on the environment. 
This is also shown by
Figure~\ref{f:dvf_zs}, where we plot the VDF for the inner subhalos ($R/R_{200}
< 0.5$; upper dotted line) and those in the cluster's periphery ($1 < R/R_{200}
< 2.5$; lower dotted line).  A power-law $dN/dv_{circ}\propto 
v_{circ}^{-4}$ is always a reasonable fit  for $v_{circ} \gsim 100\kms$.

\begin{figure}
\centering
\epsfxsize=6truecm\epsffile{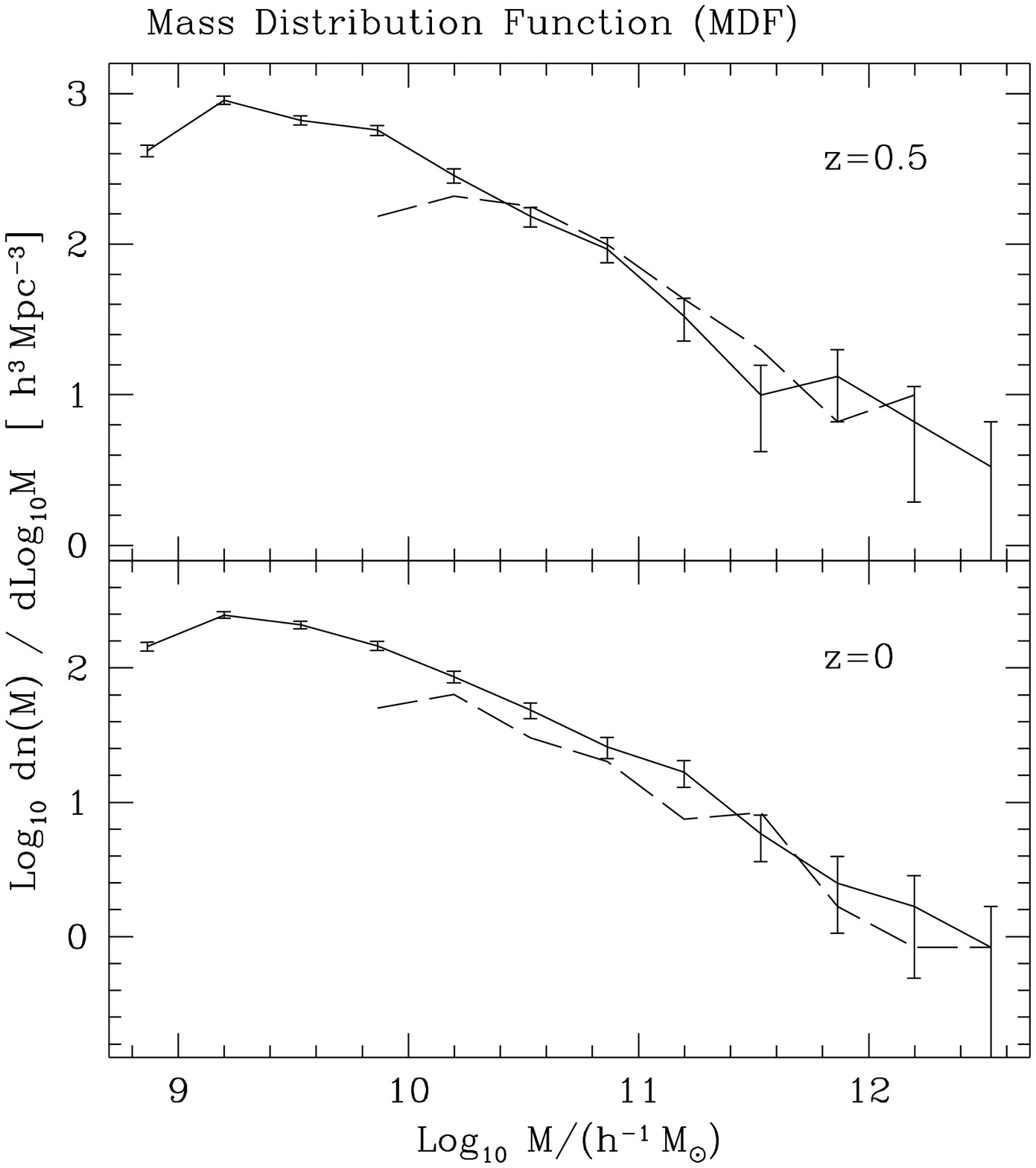}
\caption{The mass distribution function (MDF) of substructure
halos at two redhifts (lower panel: $z=0$; upper panel: $z=0.5$) for 
the two runs (LORES, dashed curve; HIRES, solid curve).  
The errorbars are (1-$\sigma$) Poisson errors on the counts.  
As for the VDF, at $z=0.5$, the curves agree well
where we expect both runs to be close to completeness 
($M\gsim 5.6\cdot 10^{10}\Msunh$).
At $z=0$, the deficiency of LORES's MDF is due to numerically enhanced mass 
loss of medium to
small mass halos in the central region (see Figure~\ref{f:dmf_coreless}).}
\label{f:dmf_2res}
\end{figure}

\begin{figure}
\centering
\begin{picture}(300, 150)   
\put(0, 60)               
{ \epsfxsize=8.4truecm \epsfysize=6.truecm       
  \epsfbox[40 300 600 720]{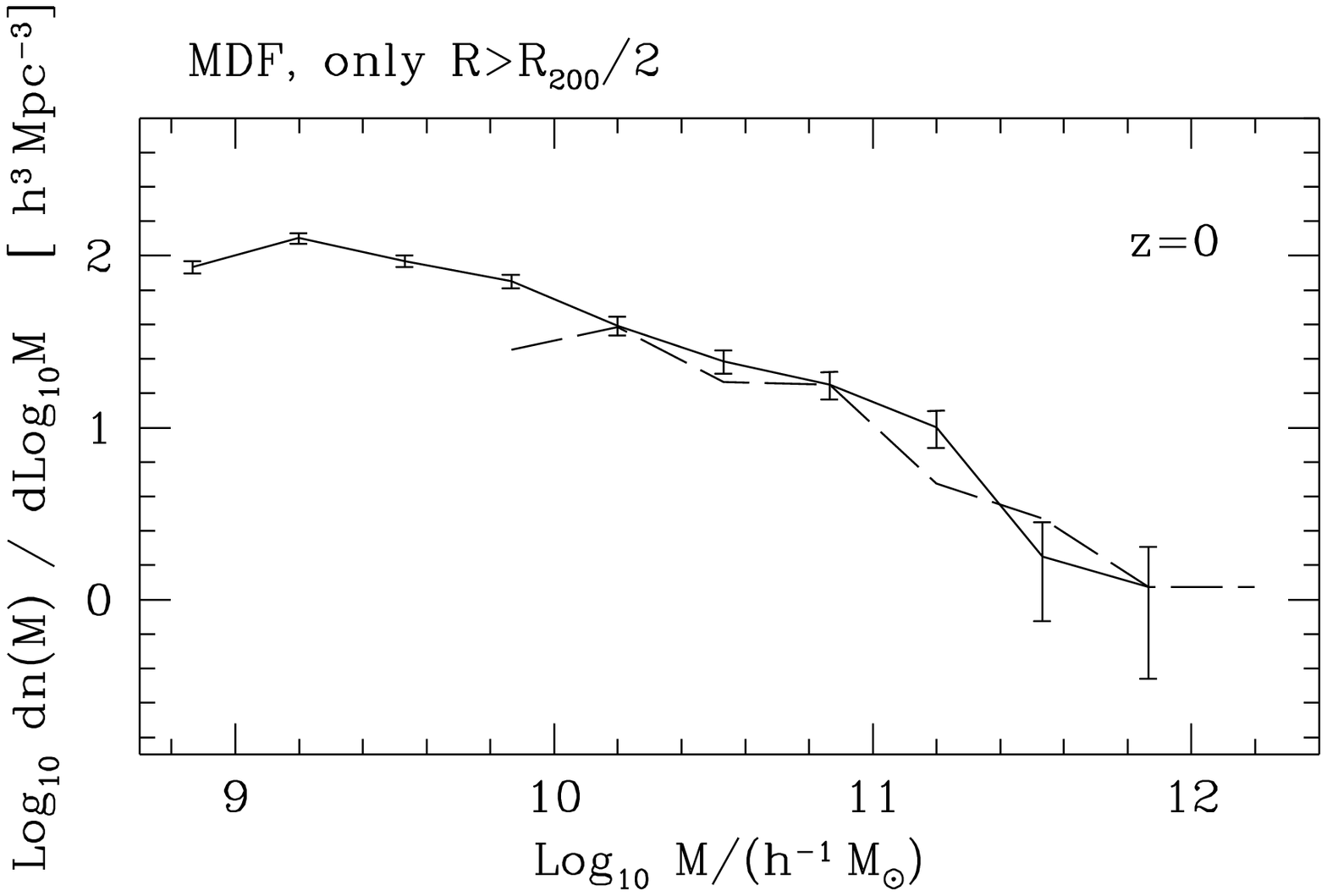}}
\end{picture}
\caption{Same as in the lower panel of Figure~\ref{f:dmf_2res}, but here the
halo samples have been constructed requiring $R > 1/2 R_{200}$. 
The curves for HIRES (solid) and LORES (dashed) are close in this case.}
\label{f:dmf_coreless}
\end{figure}

\subsection{Mass distribution function}

Knowledge of the mass distribution function (MDF) is useful to estimate the
disruptive power of halo-halo collisions in dense environments.
For example,
modelling this process is necessary to understand the origin of
intracluster light and the morphological evolution of
galaxies in clusters and groups (``galaxy harassment''; Moore \etal 1996). 
The mass and radii of substructure halos can also be probed directly using
mass-reconstruction methods based on observations of arcs and arclets in
clusters (Natarajan \etal 1998).

\begin{figure}
\centering
\epsfxsize=6truecm\epsffile{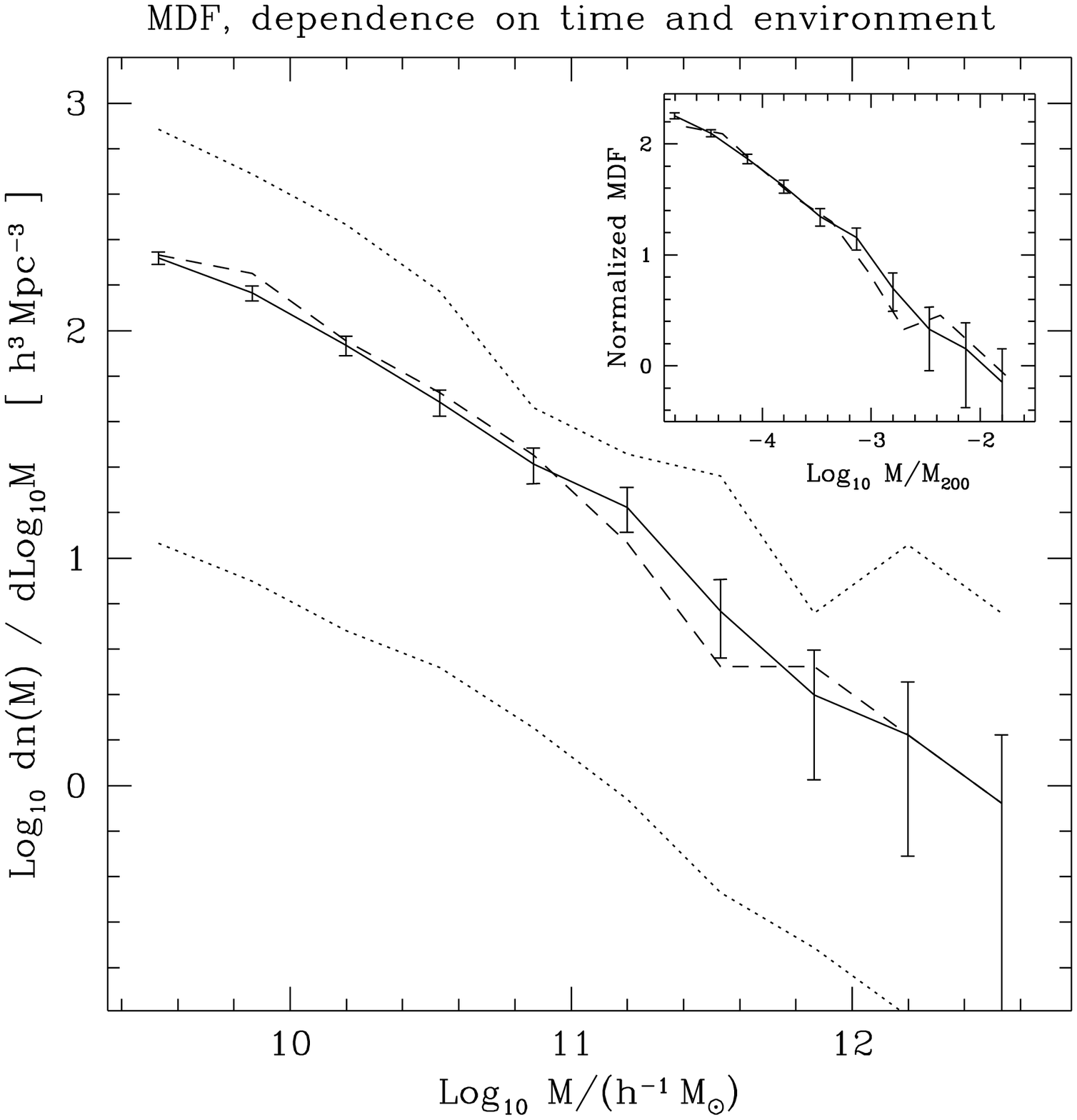}
\caption{Main panel: the MDF at $z=0$ for the cluster's halos (solid line, 
HIRES data) compared with that of the halos at $z=0.5$ contained in the 
same physical 
volume (dashed line; the curves are shown only in the range where 
they are not significantly affected by resolution).
The mass function is ``frozen in time'' 
and its shape does not depend much on the environment either, 
as it can be seen by considering only the
inner subhalos ($R/R_{200}<0.5$; upper dotted curve, main panel) or those in 
the ``periphery'' ($1 < R/R_{200} < 2.5$; lower dotted curve, main panel).
The inset shows the normalized MDFs at the two redshifts, obtained in a 
similar way as the normalized VDFs of Figure~\ref{f:dvf_zs}.}
\label{f:dmf_zs}
\end{figure}

The MDF provides information independent of 
that derived from the VDF.
The mass of a halo of given $v_{circ}$ depends on its orbital history, which
determines the extent of tidal disruption that it has suffered. This strongly
correlates with the smallest pericentric distance reached by a halo (G98)
and, to a lesser degree, with the number of orbits completed 
(since a halo loses mass very quickly the first time it passes through 
pericenter and more slowly every time it completes an orbit; 
KGKK). 
However, since halos of different masses but with the same pericenter lose
similar fractions of mass and the distribution of pericenters 
does not depend significantly on mass (G98), we still have roughly 
$M_{halo}\propto v_{circ}^3$, even for the subhalos.

\begin{figure}
\centering
\epsfxsize=6truecm
\epsffile{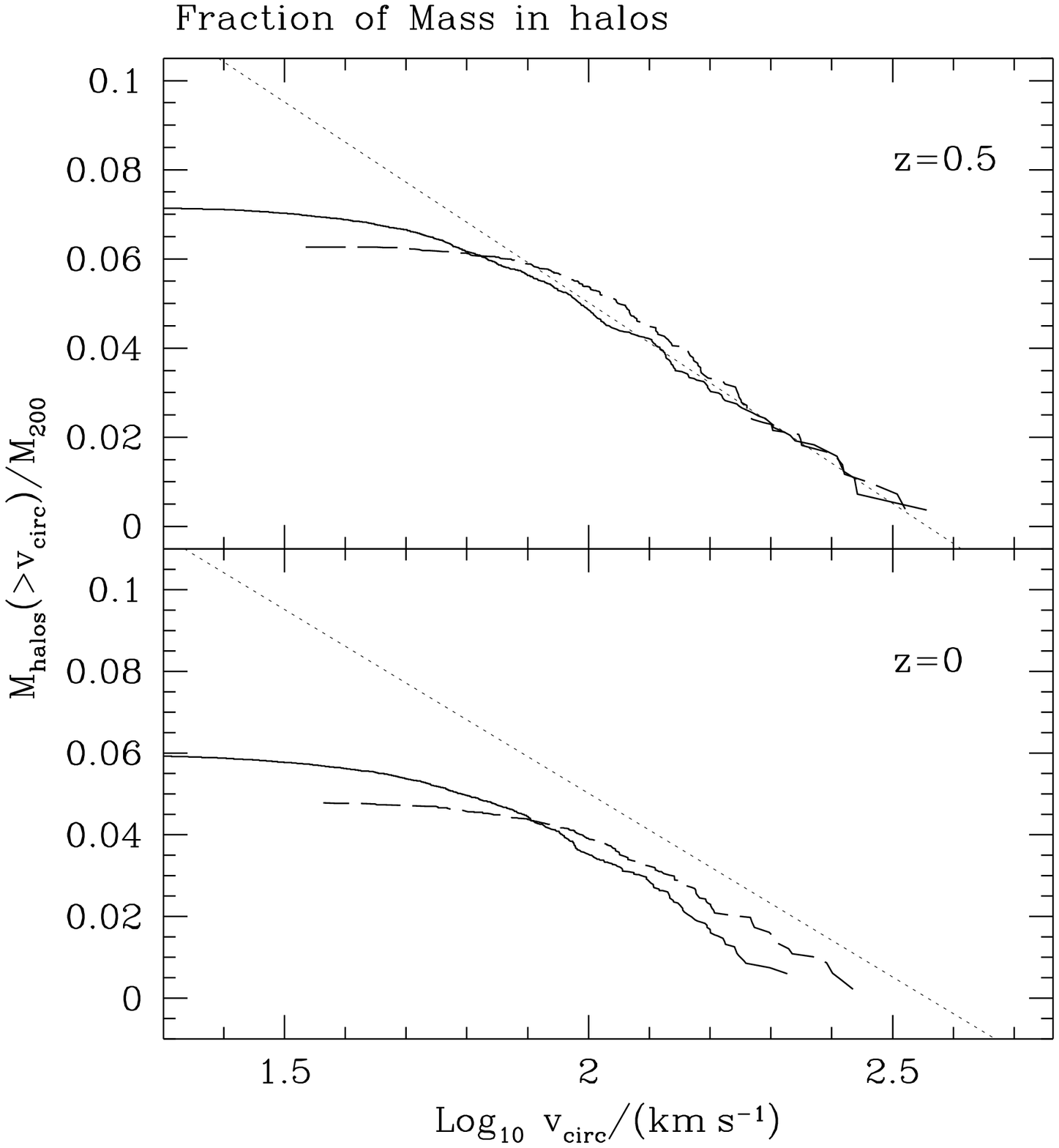}
\caption{Fraction of cluster mass bound to subhalos with 
circular velocities exceeding a given value $v_{circ}$, at 
$z=0.5$ (upper panel) and $z=0$ (lower), for HIRES
(solid line) and LORES (dashed).  
We also set an upper limit of $400\kms$ to exclude the
most massive halos whose masses have a relatively large scatter;
they contribute alone an additional $\sim 5$\%.
The light dotted line is a fit to the $z=0.5$ curves 
for $v_{circ} > 100\kms$ (the flattening of the curves at
the low-velocity end is due to the incompleteness of the samples). 
The bound mass fraction is $\propto \log(v_{circ})$, thus varies slowly
with resolution.}
\label{f:mfrac_v}
\end{figure}

Figure~\ref{f:dmf_2res} plots the MDF for the substructure halos 
(defined as the
number of halos per unit logarithmic mass interval per unit physical volume).
Results are shown for both runs at the two redshifts considered previously.  At
$z=0.5$, the agreement between HIRES and LORES is good.  The mass limit
for which both runs are close to completeness (in the circular velocity 
distribution) is $\gsim 5.6\cdot
10^{10}\Msunh$, the typical value for a halo with $v_{circ} = 100\kms$ (see
Figure~20 of G98).  At $z=0$, the MDF for the LORES cluster is $\sim 50$\%
lower than for the HIRES simulations.  This difference, as for the VDF, is
caused by numerically enhanced mass loss near the cluster's center; it 
disappears if we exclude the halos in the central region, $R/R_{200}<1/2$, 
from the counts
(Figure~\ref{f:dmf_coreless}). As discussed in the previous section, our
HIRES run is not affected significantly by artificial mass loss 
for $v_{circ}\gsim 100\kms$, i.e $M\gsim 6\cdot 10^{10}\Msunh$. 

As found for the VDF, the MDF  does not exhibit significant evolution within 
a physical volume  during the lifetime of the cluster (Figure~\ref{f:dmf_zs},
main panel). 
 At both redshifts, the MDF is close to a 
power-law $d\,n(M)/d\,M \propto M^{-2}$,
for $M_{halo}\gsim 10^{11}\Msunh$, but becomes shallower for lower masses,
$d\,n(M)/d\,M \propto M^{-1.7}$
(in the figure, we plot $dn/dLog_{10}M$). 
The main panel of Figure~\ref{f:dmf_zs} also
shows the MDF for inner and outer halos (symbols as for
Figure~\ref{f:dvf_zs}); all the curves are similar, with slopes varying 
between $-0.8$ and $-0.9$. The inset shows the normalized MDFs for 
the cluster subhalos at the two redshifts, obtained in a similar way as 
the normalized VDFs of Figure~\ref{f:dvf_zs} in the previous section.

The mass bound to substructure halos is a little more than 10\% of the
total cluster mass, with more fractional mass in halos at early rather than at
late epochs. The value varies very slowly with resolution, but it can oscillate
significantly since a large fraction of the mass belongs to the two or 
three most massive
subhalos ($v_{circ} \sim 400-500\kms$) that contribute $\approx
5$\%. If they are close to the cluster's center or if they possess their own
substructure, the masses measured by SKID are quite sensitive to the linking
length used and can differ significantly from estimates obtained adopting the
spherical overdensity method (see G98), sometimes by a factor of
2. For instance, using the spherical overdensity method for LORES we find a 
global fraction of mass in subhalos of $\sim 13$\% (G98), while using 
SKID's masses this value reduces to $\sim 11$\%.
If we exclude the largest substructure halos, different mass estimates yield
halo mass fractions differing only by about 1\%.

Figure~\ref{f:mfrac_v} plots the fraction of cluster mass that is in subhalos
with circular velocity above a given value $v_{circ}$, but below $400\kms$ The
difference between the two runs never exceeds 1\%. Also, between $z=0.5$ and
$z=0$, the mass fraction decreases by only $\sim 1$\% (note that the mass
fraction is computed relative to the virial mass, $M_{200}$, at the respective
epoch, namely $1.7$ and $2.15\times 10^{14}\Msunh$). Where the halo samples 
are close to completeness, the results are well represented by the function
$M_{halos}(v_{circ}>\bar{v}_{circ})/M_{200} \simeq -0.09
\log_{10}[\bar{v}_{circ}/(\kms)] + \beta$, with $\beta = 0.23$-$0.22$. 
 The mass
fraction varies roughly linearly with clustercentric distance (G98,
Figure~11).

\begin{figure}
\begin{picture}(300, 460)  
\put(0, 290)               
{\epsfxsize=8.4truecm \epsfysize=6.truecm       
\epsfbox[40 300 600 720]{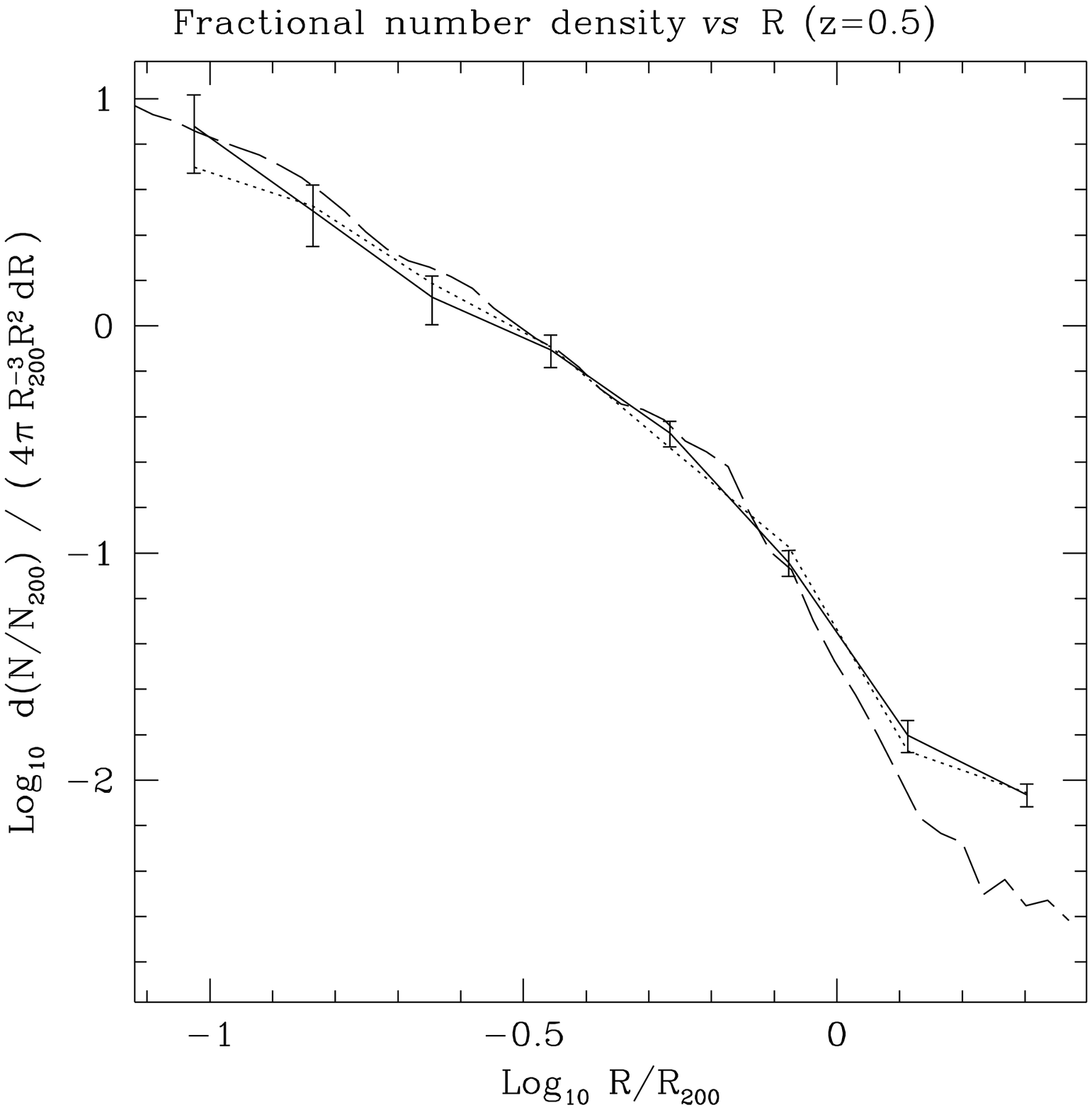}}
\put(0, 60) 
{\epsfxsize=8.4truecm \epsfysize=6.truecm       
\epsfbox[40 300 600 720]{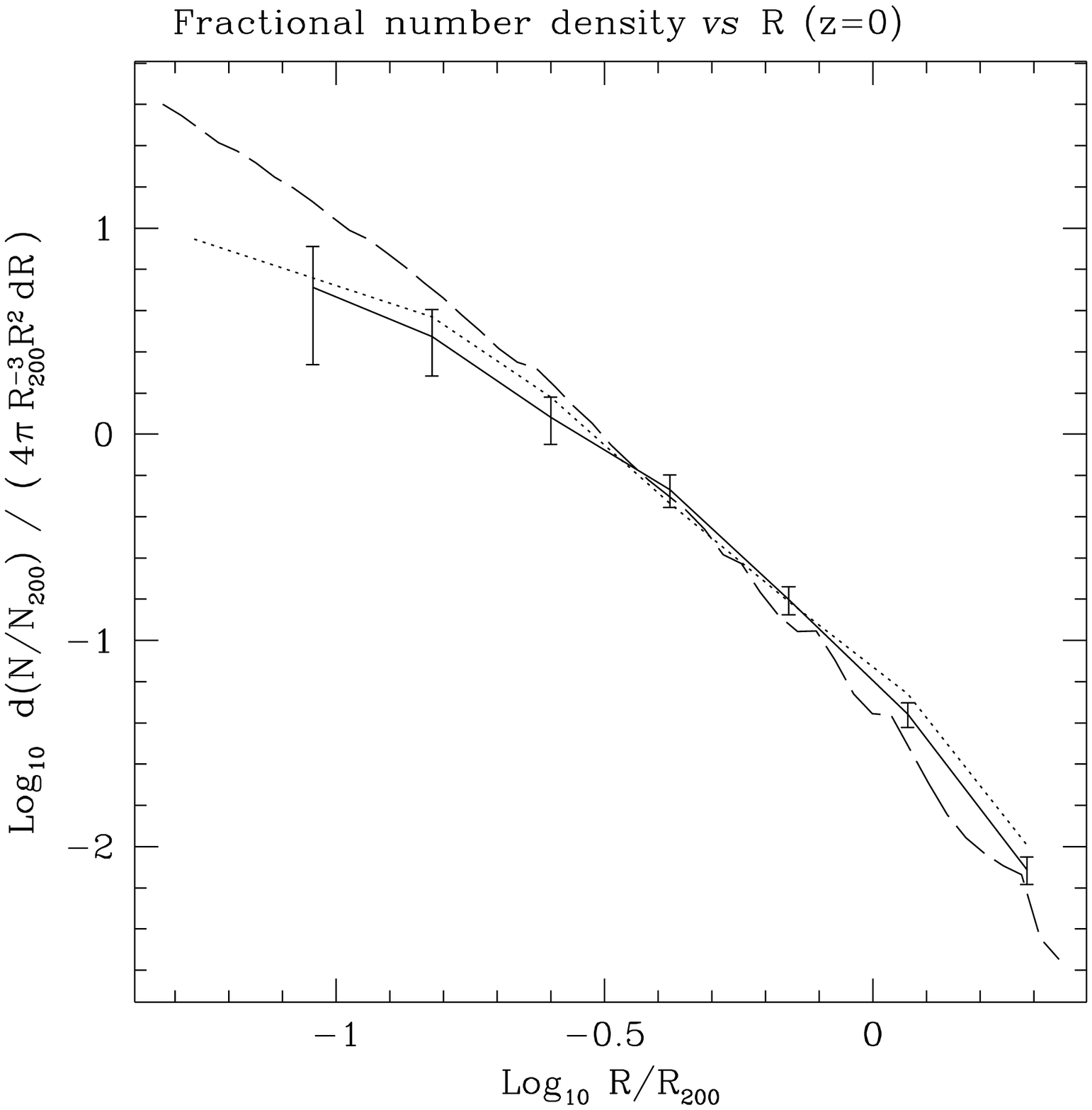}}
\end{picture}
\caption{Fractional number density of halos and particles as a function of
clustercentric distance $R$, at $z=0.5$ (upper panel) and $z=0$ (lower panel);
lengths are in units of the virial radius $R_{200}$ at the respective epoch; 
$N_{200}$ is the number of halos or particles within $R_{200}$.
In the top panel,
the solid curve (with 1-$\sigma$ Poisson
errobars) is for HIRES's subhalos with $v_{circ}> 80\kms$; 
the dotted curve is LORES's equivalent.
The dashed line is the particle (i.e. the mass) profile.
The subhalos are always antibiased with respect to the mass, 
although at $z=0.5$ their profile is similar to that of the mass
within the cluster.}
\label{f:ndens_vs_P}
\end{figure}

\subsection{The spatial distribution of substructure halos}

In this section, we examine the bias existing between 
the spatial distribution of particles (i.e., the mass)
and that of the subhalos. 
%
Here, the bias is defined as the ratio $b(R)\equiv n_{hal}(R)/
n_{part}(R)$, where $n_{hal}(R)$ and $n_{part}(R)$ are the number density 
profiles of subhalos and particles as functions of clustercentric distance
$R$. For convenience we normalize all quantities to the cluster parameters,
i.e. we measure lengths in units of the virial radius, $R_{200}$, 
and $N_{hal}$ and $N_{part}$  as fractions of the total 
numbers of subhalos and particles within $R_{200}$. The halo distribution will
be (positively) biased if $b(R)$ increases as $R$ decreases, anti-biased in 
the opposite case. 
In G98 we found that the halos are anti-biased;
Col\'{\i}n \etal (1999) also find anti-bias for dark matter halos
in large volume simulations of four cosmological models 
(they measure the bias defined by the ratio between the 2-point 
correlation functions of mass and dark matter halos, as it is
customary for large scale structure studies). 
To study this problem it is particularly important 
to pay attention to the limitations imposed by the finite resolution, since
overmerging --- preferentially erasing subhalos in the central regions of the 
cluster --- always introduces 
anti-bias in the subhalo distribution (or enhances it). 

We compare the {\sl normalized} (or fractional) number density profiles 
of halos and particles 
as a function of clustercentric distance $R$ in Figure~\ref{f:ndens_vs_P}, 
for the redshifts $z=0.5$ and $z=0$ (top and bottom panel respectively).
We plot the numbers of halos with 
$v_{circ}>v_{circ}^{LIM}$ in spherical 
shells of radius $R$ divided by the volume of the shell 
(in units of $R_{200}$), divided by 
the total number ($N_{200}$) of cluster subhalos above the circular velocity 
limit (the ``cluster'' subhalos are those contained in $R_{200}$ at 
the respective epoch;
this yields 188 and 219 subhalos if $v_{circ}^{LIM}=80\kms$, and 
104 and 110 if $v_{circ}^{LIM}=100\kms$ for HIRES, while for LORES 
the corresponding numbers are 165 and 158, and 116 and 100).  
For the mass, we plot the particle number density profile divided by the
number of particles within $R_{200}$. 
In each panel, the {\sl long dashed} curves are the normalized mass 
profiles; the two other curves are for the halos of HIRES 
({\sl solid with errobars}) and LORES ({\sl dotted}). For $z=0.5$, we 
set $v_{circ}^{LIM}=80\kms$, which yield good 
statistics; for $z=0$ we use  $v_{circ}^{LIM}=100\kms$, which has similarly
good statistics. 
Varying the value of $v_{circ}^{LIM}$ does not make any 
difference, as long as there are enough halos to obtain significant measures.

\begin{figure}
\centering
\epsfxsize=6truecm
\epsffile{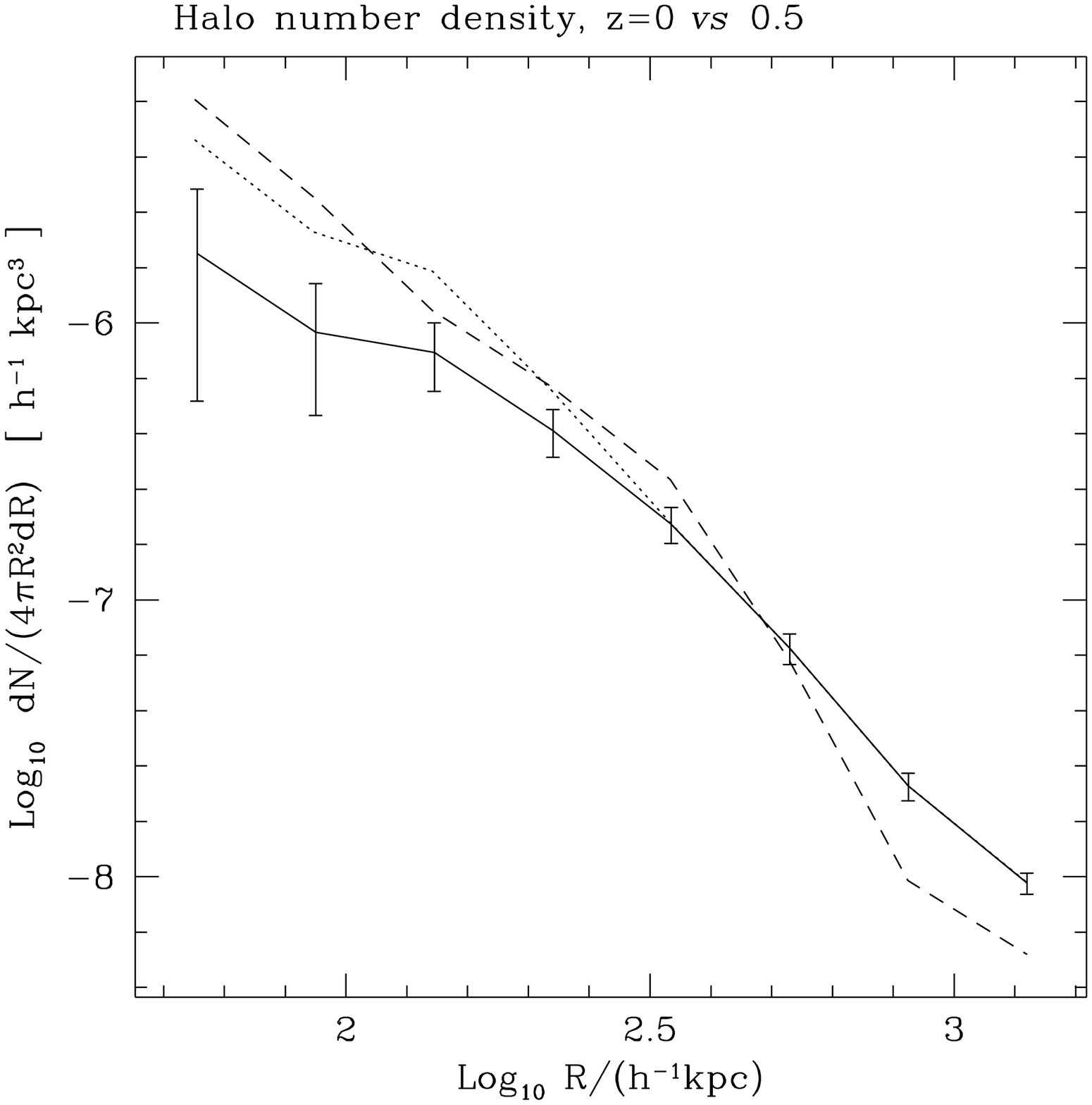}
\caption{The number density profiles in physical (proper) volume using
halos with $v_{circ}> v_{circ}^{LIM}\equiv 80\kms$ at $z=0$ (solid line) 
and $z=0.5$ (dashed) .
The halo number density decreases in the central region ($R
\lsim R_{200}/3$). The dotted line shows the effect of having a variable
$v_{circ}^{LIM}$ that drops by $\sim 20$\% for $R < R_{200}/4$ (precisely, 
by 15\% in the radial range $0.125<R/R_{200}<0.25$ and by 25\% for
$R/R_{200}<0.125$).}
\label{f:ndens_zs}
\end{figure}

Within the cluster, at $z=0.5$, the halos have a number density 
profile similar to that of the mass, but
the latter profile declines more steeply than the former approaching the 
virial radius and beyond (i.e., $b(R<R_{200}) < b(R>R_{200}$).
 This means that the population of halos
(for a circular velocity limited sample)
in the volume encompassing the cluster is globally anti-biased -- with 
proportionally fewer halos where the particles are more clustered.
At $z=0$, there is also anti-bias, the mass being
clearly more concentrated than the halos; 
for $R\sim 10$\% of $R_{200}$, there is a factor of 2 
of difference between the two profiles (and is even larger if we require
that the two profiles overlap beyond $R_{200}$). 
At both redshifts, there are no significant differences between the curves for
HIRES and LORES.
Therefore the halo number density profiles measured from this simulation
should not be significantly affected by residual overmerging (see also
\S~6). 

Figure~\ref{f:ndens_zs} compares directly the halo number density profiles 
at $z=0.5$ and $z=0$ in the physical (proper) volume centered on the cluster
(for both samples, we set $v_{circ}^{LIM}\equiv 80\kms$). 
The halos are clearly more concentrated in the volume of the newly formed
cluster. Infalling material subsequently enhances the halo number density 
in the outer region, $R\gsim 600\kpch$ (the virial radius of the cluster at 
$z=0.5$), while tidal disruption in the inner region reduces the 
number density of halos with circular velocities above the limit. 
%
It is interesting to note that allowing for a $\sim 20$\% decrease of a halo's
$v_{circ}$ between $z=0.5$ and $z=0$ (because of central mass loss, 
as observed in the VDF analysis in
\S~5.3) is sufficient to recover most of the
central steepness of the $z=0.5$ profile; this is shown by the dotted line
in the Figure. This indicates that, once the cluster is in place, tidal mass 
loss alone (instead of total halo evaporation) 
causes the measured decrement of the subhalo number density towards the 
cluster's center.

Apart from tidal mass loss, mergers between halos during the cluster assembly 
(at $z>0.5$) are probably a major cause of the anti-bias between 
halos and particles. 
At $z=0.5$, the uniform 
anti-bias of the cluster subhalos (with respect to the `field'; as seen in
Figure~\ref{f:ndens_vs_P}, upper panel) may 
reflect the fact that the cluster has only recently being assembled through 
the merging of group-sized halos of similar masses, so that 
mergers and tides have affected its subhalo population more uniformly. 
Note that the galaxies that would reside in the subhalos could have a 
different bias, if their circular velocities are not directly 
affected by the central mass loss of their dark halos.
As a final word of caution, we point out that the issue of bias would be
better addressed with a statistical sample of clusters and a more
representative sample of field halos. This approach  
requires a trade-off between statistics and resolution but considerable 
progress has recently been made  
(Col\'{\i}n \etal 1999; see also Kauffmann \etal 
1999a,b, Diaferio \etal 1999, Benson \etal 2000).

\subsection{Velocities of halos and dark particles}

\begin{figure}
\begin{picture}(300, 460)  
\put(0, 290)               
{\epsfxsize=8.4truecm \epsfysize=6.truecm       
\epsfbox[40 300 600 720]{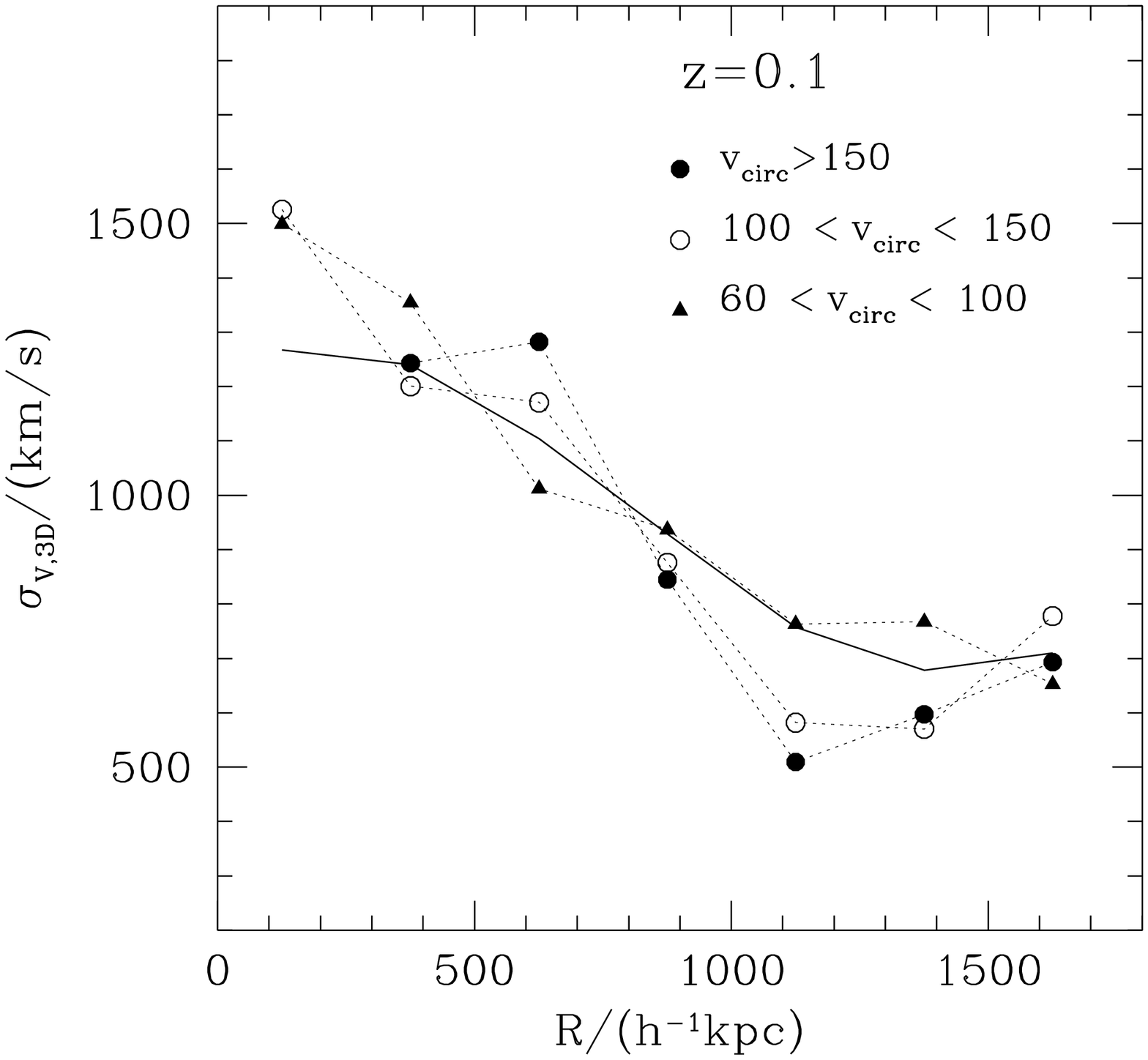}}
\put(0, 60) 
{\epsfxsize=8.4truecm \epsfysize=6.truecm       
\epsfbox[40 300 600 720]{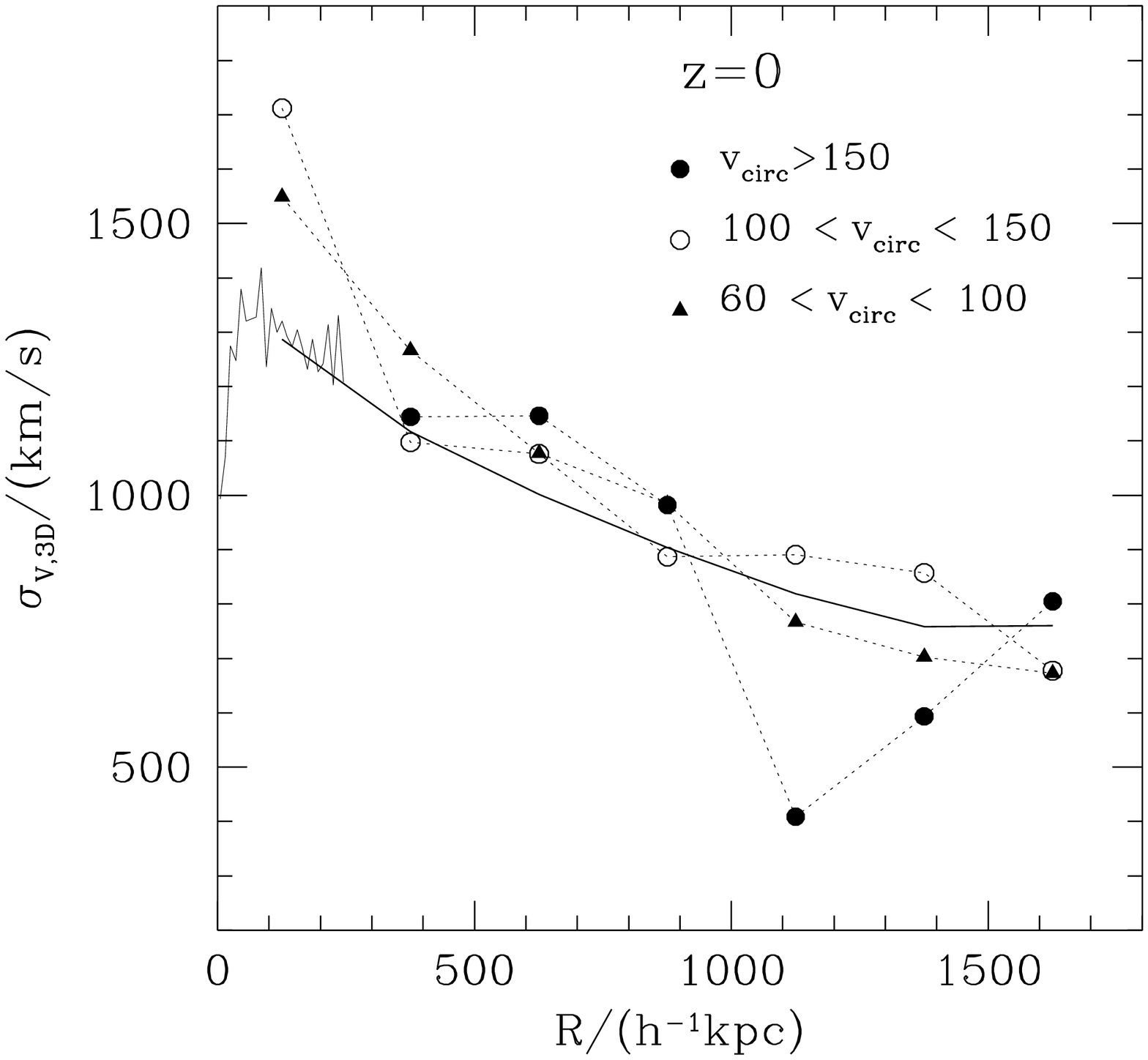}}
\end{picture}
\caption{3D velocity dispersion profiles for the dark matter (solid curve)
and for three samples of halos (the points on the dotted curves) 
selected according to their circular velocities as outlined on the figure
($v_{circ}$ in $\kms$). The data are for HIRES at the redshifts $z=0.1$ and
$z=0$; $R_{200}$ is respectively $0.88\Mpch$ and $0.975\Mpch$.  
(The curve for $v_{circ}>150$ does not extend to the 
innermost bin, since the latter contains only two such halos - the ``cD'' 
has been excluded (it is always at rest since it simply the central smooth
cluster potential; the profile for the dark matter using fine bins near the cluster's 
center is shown by the light solid line in the lower 
panel). The halos in the central region ($R\lsim 300\kpch$) 
are ``hotter'' than the DM, with $b_v\sim 1.2$-$1.3$.}
\label{f:v3d}
\end{figure}

\begin{figure}
\centering
\epsfxsize=6truecm
\epsffile{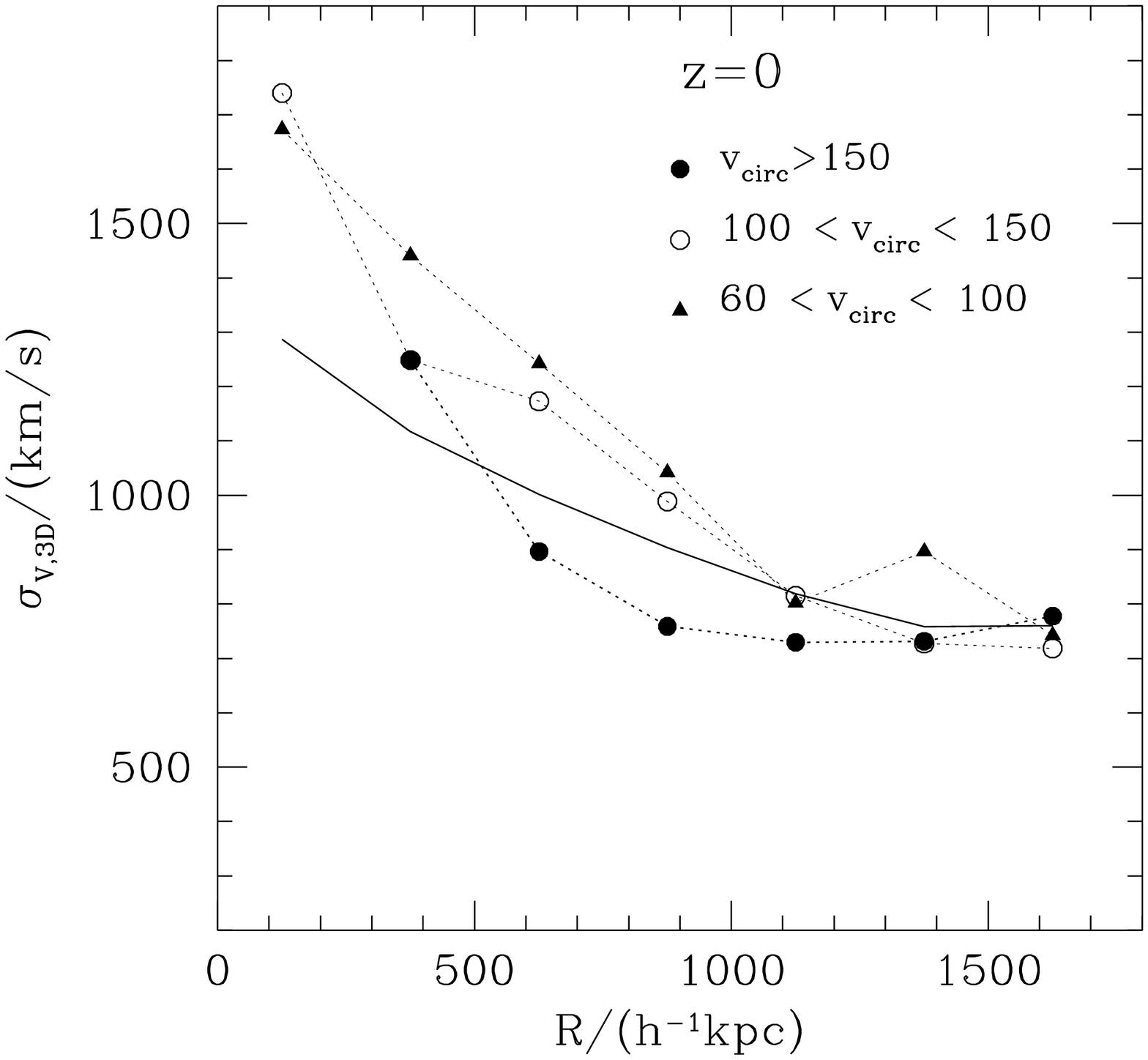}
\caption{Same as in Figure~\ref{f:v3d} but using data from LORES at $z=0$.
The positive bias of the halos in the central region is more marked than 
for HIRES.}
\label{f:v3d_2kpc}
\end{figure}

The issue of velocity bias --- the difference between galaxy and dark matter 
(DM) velocities --- was first addressed by Carlberg \& Couchmann (1989) 
as a means of
 reconciling low measures of $\Omega$ from the observed
velocities of galaxies in clusters. There are two types of velocity bias 
considered in the literature, the 
one-point velocity bias ($b_v=\sigma_{V,halos}/\sigma_{V,DM}$) 
which compares the velocity dispersions of galaxies 
 and dark matter particles (e.g. Carlberg \& Dubinski 1991, Carlberg 1994) 
and the two-point or pairwise velocity bias 
($b_{v,12}$) which
compares the relative velocity dispersions in pairs of objects (e.g.
Couchmann \& Carlberg 1992).  Here we 
consider the first type. A recent paper (Col\'{\i}n, Klypin \& Kravtsov
 1999; CKK hereafter) focussed on the different forms of
velocity bias for numerical simulations of the $\Lambda$CDM cosmology.
The situation is still rather confusing. CKK find that the 
dark matter halos within clusters are positively biased, 
with $b_v \sim 1.2$-$1.3$
(similar results were also found by Diaferio \etal 1998 and  Okamoto \& Habe
1999), while KGKK find $b_v\lsim 1$ for a similar set of 
data (see their Figure~15). In G98, we did not detect velocity bias.
The difference may be partly due to the fact that 
G98 and KGKK measure the global $b_v$
using the whole sample of cluster subhalos, while CKK consider the radial
profile $b_v(R)$.  We re-examine the issue here in more detail. 
Note that the comparison between the 
spatial distribution of
the halos and that of the mass considered in the previous section, does not
provide sufficient information, since the scaled profiles of differently biased
tracers may not differ much (Carlberg 1994).

We compare the 3D velocity dispersion ($\sigma_{V,3D}$) 
profiles of halos and dark matter for run HIRES in Figure~\ref{f:v3d} (the averages are
taken in spherical shells of width $250\kpch$; at the center, 
$\sigma_{V,3D}$ for the 
DM drops sharply from $\sim 1300\kms$ at $R\sim 50\kpch$ to $1000\kms$ at 
$R\sim 10\kpch$, if radial bins of width $10\kpch$ are used, as it is 
shown by the light solid curve in the lower panel of the figure).
We consider the cluster at $z=0$ (lower panel) and also the output 
at $z=0.1$, separated in time from the final epoch 
by less than one typical subhalo orbital period; we also subdivide the 
halos in three subsamples according to the values of their circular
velocities (as listed in the figure). In this way, we should be able
 to single out 
significant differences between the (3D) velocities of halos and dark matter from
statistical flukes, partly overcoming the drawback of dealing with only one
cluster.  The samples with $v_{circ}>150\kms$ have the worst 
statistics, especially for $R>R_{200}$ where they provide $\sim 5$ halos
per bin.

With quite large oscillations, the $\sigma_{V,3D}$ of 
the halos is consistent with that of the dark matter for $R \gsim 300\kpch$.
Within the cluster, the data show a significant (positive) bias in the 
innermost bin, with $b_v=1.2$-$1.3$. 
The maximum amplitude of bias in our data is similar to that reported by CKK
but the signal is globally weaker (CKK have $b_v\simeq1.2$ 
at $R/R_{200}\simeq0.7$, that is $\gsim 600\kpch$ for the cluster studied 
here).
We also show the velocity dispersion profiles of halos and 
particles for run LORES, in Figure~\ref{f:v3d_2kpc}. The trend of
increasing $b_v$ with decreasing $R$ is more apparent  in this case.
Here, the sample with the lowest circular velocities is largely incomplete
due to limited resolution; interestingly, it exhibits a more marked bias
than the other samples. 
It is also interesting to note that there are no obvious biases
in the distribution of the orbital parameters of the halos and 
the particles (we have computed the orbits approximating the cluster
potential to a static spherical potential obtained from its density
profile, as in G98, \S~4.6).

When we average over the virial volume of the cluster, 
the velocity bias of the halos in the central region
has a small impact; in the three cases considered previously, the
global bias within $R_{200}$ is always less than $1.1$ (although it 
remains positive). Also considering only  massive halos
($v_{circ}>200\kms$) does not yield any significant bias.
In this case, anti-bias would be expected if dynamical friction were
efficient in slowing down such halos, but the friction time 
for subhalos within a cluster is almost always larger than the 
Hubble time (see Colpi, Mayer \& Governato 1999 for a recent analysis).
Even restricting the samples to half the virial radius does not make a 
significant difference (in one case, for HIRES at $z=0.1$, there
is anti-bias, but the other two yield positive or no bias, unless 
the ``cD$\equiv$global potential'' is included). 

In conclusion, we find evidence of positive velocity
bias in the central region of the cluster but the signal is much weaker than
CKK's detection. It seems that
limited resolution enhances the signal; overmerging accentuates 
the impact of physical processes like
mergers and total halo disruption that could lead to positive velocity bias.
A larger set of high-resolution simulations would be welcome to
address these issues with better statistics. 

\section{Survival of substructure halos}
\begin{table}
\begin{tabular}{|l|c|c|c|c|}                  \hline \hline
$v_{circ,\,min}$  &  $N_{halos}$  &  $p_{traced}$  & $p_{cD}$ \\ 
\hline 
 $230\kms$ & 15 & 90\% & $\sim 10$\% \\
 $100\kms$ & 120 & 100\% & 0\% \\
 $50\kms$ & 550 & 82\% & 0\% \\
\hline
\end{tabular}
\caption{Results of tracing the cluster's progenitor halos at $z=1$ to the 
substructure halos at the final time ({\sl descendents}), for three ranges 
of mass of the progenitors (corresponding 
to more than $10,000$, between $10,000$ and $1,000$, and between $1,000$ and
100 particles). The first and second columns report the 
minimum circular  velocity of the progenitors in each mass range and 
their number; the third and fourth
columns give, respectively, the percentage of progenitors in each  range 
with a descendent at $z=0$ other than the ``cD'' object, and 
the percentage of progenitors that merge with the ``cD''. 
(These values are obtained
requiring for the descendents a minimum number of traced particles 
$N_{p,min}$ equal to 1\% of the particles in the progenitor and not less 
than 4; there is no difference 
if we set $N_{p,min}$ to be at least 16).} 
\label{t:tab1}
\end{table}

\begin{table}
\begin{tabular}{|l|c|c|c|c|}                  \hline  \hline
$v_{circ,\,min}$  &  $N_{halos}$  &  $p_{traced}$  & $p_{cD}$ \\ 
\hline 
 $300\kms$ & 23 & 40\% (40) & 60\% (60) \\
 $150\kms$ & 180 & 90\% (90) & 10\% (7) \\
 $70\kms$ & 1070 & 70\% (61) & 3\% (0.1) \\
\hline
\end{tabular}
\caption{Results of tracing the cluster's progenitor halos at $z=3$ to the 
substructure halos at the final time. The mass ranges and
the quantities in the columns are the same as in table~1
(in the last two columns, the values in 
brackets give the percentages obtained requiring $N_{p,min}$ to be at least 
16).}
\label{t:tab2}
\end{table}

As seen in \S~5, over the lifetime of the cluster 
the abundance of substructure does not appear to change significantly;
however substructure is erased also at earlier redshifts. 
To examine this issue, we
have identified the dark matter halos that are progenitors of the cluster
at $z=1$ and $z=3$ (using the public friends-of-friends code available
from http://www--hpcc.astro.washington.edu/tools, run with a linking length 
of $0.2$ times the mean interparticle separation). We define as progenitors 
those halos 
that contribute particles to the virial volume of the cluster 
at the final epoch. 
We have then ``traced'' them
to the substructure halos identified at $z=0$, by comparing the indexes of 
their particles. For the $z=0$ subhalos
we use only the `core' particles, $i.e.$ those that are
separated from the halo centers by a distance less than the value of 
$r_{peak}$ - this (conservative) approach avoids ambiguities when 
there is a hierarchy of substructure. 
 For the central object, we adopt a radius $r_{cD}=25\kpch$, a value 
that should 
single out the progenitor halos that merge to form the ``cD'' core 
against those that only supply particles to its halo through tidal stripping.

We accept descendents only if they have
a number of traced particles larger than 1\% of the particles of the
progenitors and not less than a minimum number, $N_{p,min}$, which we varied
between 4 and 16 (the number of particles traced
can be small for small halos since we only consider their `core' particles).
In our approach, a progenitor with no descendent subhalos at $z=0$ should have
been completely disrupted with its particles 
lost in the diffuse particle background of the cluster or, possibly, in the 
outer parts of large subhalos (visual inspection of 10 cases confirms
this).  Such halos may survive if the 
numerical resolution were increased further, so that the fraction of 
progenitors with no descendents is an estimate of the importance  
of residual numerical effects.
Cases in which two or more progenitors have a common 
descendent, i.e. the progenitors' particles contribute to the descendent's 
core, point to true mergers.

Table~1 reports the results of the tracing 
for the $z=1$ progenitors in three mass ranges.
Table~2 is the analog for $z=3$. In the first column, 
we show the  minimum value of the circular velocity for the 
progenitors in each mass range. 
All the $z=1$ progenitors  with $v_{circ}>100\kms$ 
have a descendent at $z=0$; the tracing is still essentially complete
at $v_{circ}\gsim 70\kms$ ($p_{traced}=98$\%) and deteriorates only 
for $v_{circ}$ approaching $50\kms$ ($p_{traced}\simeq 80$\%). 
The central object accretes only another halo (of $v_{circ}\sim 300\kms$).
These results are stable against varying $N_{min}$.
The loss of halos
from $z\sim 1$ to the present appears negligible for $v_{circ} > 100\kms$.

The most prominent feature of the tracing of the $z=3$ progenitors is the large
fraction of massive halos that merge with the central object:
about 25 progenitors with  $v_{circ} > 150$, a dozen for
$v_{circ}>300\kms$.  There is virtually no contribution to
the central object's (core) mass from progenitors with  $v_{circ} < 150$.
Essentially all the progenitors with $v_{circ}>150\kms$ have a partner at
$z=0$; the tracing starts deteriorating at $v_{circ}\lsim 100\kms$ where the
percentage of ``childless'' progenitors is $18$\%, independent of 
$N_{min}$ in the range 4-16. 
Successful tracing with the ``cD'' might still hide overmerging, but 
it seems safe to consider the mergers of the massive halos as real,
since they occur before the cluster formation and involve objects 
of comparable masses (for which the dynamical friction time is small). 
Also the distribution of their particles in the $z=0$
cluster is well concentrated within the cD.   
We have examined the distribution at the final epoch of the 
particles in the cores of the $z=3$ progenitors with $v_{circ} >200\kms$ 
that merge with the cD (we use $r_{core}=20\kpc$ comoving); 
about 80\% of such particles 
end up within $50\kpch$ from the cluster's center at $z=0$.
Very likely further increases of 
resolution, and the inclusion of baryons with cooling, will reveal 
substructure within a cD's radius; e.g. the substructure already present
within the progenitors at $z=3$ could survive and be 
found within the 
cD. However, 
these subhalos would lose a very large fraction of their masses 
and should have small circular velocities and low mass-to-light ratios.

In conclusion, overmerging effects in this simulation 
appear negligible for large and intermediate halos;
loss of substructure may still
affect halos starting from $v_{circ} \sim 100\kms$ at the $\sim$20\% level.
This does not seem to be important for most of the statistical properties 
of the substructure considered in the previous section, in view of the strict 
similarities of the results for LORES and HIRES (except possibly for the
velocity bias).   

\section{Summary and Conclusions}

We use ``N-body'' simulations to follow the collisionless
evolution of the dark matter component of a rich galaxy cluster.
Increasing the force and mass resolution (Figure~\ref{f:2maps}) by an
order of magnitude has a dramatic effect on the abundance of substructure
and allows us to assess the biases introduced by low resolution studies.
We find the following main results:

\begin{itemize}

\item{} We compare the density profile of the cluster halo simulated
with $2\times 10^4$, $6\times 10^5$ and $4\times 10^6$ particles
within the virial radius. At our highest resolution the force
softening is $0.5\kpch$ ($0.05$\% of the virial radius) and the
profile is well fit by $ \rho(r) \propto
[(r/r_s)^{1.5}(1+(r/r_s)^{1.5})]^{-1}$.
This agrees with the results of Moore \etal
(1999b) for lower resolution simulations, and suggests that we may
have converged on the asymptotic central slope. The best fit NFW
profile to the data has residuals of order 30\%.
(In a recent paper, Jing \& Suto, 2000, find shallower central
cusps ($\rho(r)\propto r^{-1.1}$) for four high resolution 
simulations of clusters in the $\Lambda$CDM model; the 
reasons of the  disagreement are not yet clear, see \S~4).  

\item{} For the first time we can confidently resolve the cluster 
density profile beyond $\approx 0.2\% R_{vir} = 2\kpch$.
Between $2\kpch$ and $40\kpch$ the density profile has a central
cusp with a slope $\rho(r) \propto r^{-1.5}\pm0.05$.  This 
disagrees with the central mass profile of CL0024+1654 recovered
from a strong lensing analysis which indicate a constant density
core over the same radial range (Tyson \etal 1998).

\item{} The density profiles of substructure halos has not yet
``converged''. Lower resolution studies found that halos within clusters
had steeper density profiles than field halos -- this may be due to
particle-halo heating of substructure halos.

\item{} The distribution function of substructure circular velocities
(VDF) is essentially invariant over the lifetime of the cluster
($\sim5$ billion years) and is quite close to a power-law
$dn(v_{circ})/dv_{circ} \propto v^{-4}$, with little dependence on
environment (\S~5.1).  Clusters simulated with $\sim 10^6$ particles
can reliably measure this function for halos with circular velocity
above $100\kms$. (Note that the cluster forms early, at $z\gsim 0.5$,
with no major mergers since; the evolution of the properties
of the substructure may be different in objects with more turbulent recent
histories.) 

\item{} The distribution of halo masses (MDF) shows the same
``invariance properties'' as the VDF --virtually no evolution and
little dependence on environment-- and is close to a power-law $dn/dM
\propto M^{-2}$ for halo masses $\gsim 10^{11}\Msunh$ 
(a fraction $5\times 10^{-4}$ of the cluster's virial mass); 
tidally driven mass loss changes the MDF
self-similarly (\S~5.2).  Clusters simulated with
$\sim 10^6$ particles can reliably measure this function for halos
with masses above $6\times10^{11}\Msunh$.

\item{} The substructure halos of circular velocity limited samples 
are spatially anti-biased with respect to the underlying
mass distribution (see \S~5.3 for definitions).
 At an early epoch ($z=0.5$), the anti-bias is uniform
within the cluster (halo and particles have number density profiles of
similar shape); at the final time, the anti-bias of the halos increases
approaching the cluster's center. Halo-halo collisions and tidal mass
loss (which decrease a halo's $v_{circ}$, on average by $\sim 20$\% over 
5 billion years) are the likely source of the anti-bias (\S~5.3). 

\item{} The 3D velocity dispersions of halos and dark matter
particles are generally consistent within and around the cluster,
except in the central region ($R< 0.3R_{200}= 300\kpch$) where the
halos are ``hotter'' than the particles with a (positive) velocity
bias $b_v\equiv\sigma_{v,halos}/\sigma_{v,DM}=1.2$-$1.3$). The sign
and maximum amplitude of the bias is similar to that reported by
Col\'{\i}n \etal (1999), but the radial range over which $b_v>1$ is
much more limited (their detection extends to $R\sim 0.8R_{200}$).
The magnitude of the bias appears to be enhanced by limited resolution
since ``overmerging'' erases preferentially the central cluster halos. 
When we average over $R_{200}$, the velocity dispersions of halos and
particles do not differ by more than 10\%. 

\item{} Following the cluster progenitors from high redshift to the
present shows that, at the current resolution, overmerging should be
globally unimportant for subhalos with $v_{circ} \gsim 100\kms$,
although a fraction of high-redshift halos may be lost for $v_{circ}$
close to $100\kms$. All the cluster progenitors with $v_{circ}\gsim
70\kms$ at $z=1$ have a descendant among the subhalos at $z=0$; about
$20$\% of the progenitors at $z=3$ with $v_{circ}\sim 100\kms$ may be
artificially disrupted. The ``cD'' object at the cluster's center is
assembled at high redshift ($z\sim 3$ to 1) through the
merging/accretion of a dozen halos with $v_{circ}\lsim 300\kms$; it
also accretes one large halo within the cluster (at $z<0.5$).

\end{itemize}

It is now clear that the hierarchical model for structure formation can
naturally form massive dark matter halos that contain a wealth of substructure 
resembling observed galaxy clusters;
high resolution numerical simulations
allow us to obtain robust statistical measures of the properties of the 
substructure and make detailed predictions.
Since substructure halos are expected to host galaxies in a cluster and galaxy
satellites in a galaxy, the comparison with the observed 
distribution of those objects 
can produce important tests for the cosmological models.
For example, whereas the MDF obtained from cluster 
simulations in the CDM universe easily 
reproduces that derived from the observed
circular velocity function of cluster galaxies (Moore \etal 1999a),
the predicted abundance of substructure of galactic halos is not matched
by the observed population of satellites of the Milky Way and Andromeda 
(Klypin \etal 1999b, Moore \etal 1999a). 

\section*{Acknowledgments}
SG is a Marie Curie Fellow of the TMR program of the 
European Union (4-th framework, grant n. ERBFMBICT972103). 
We are grateful to Carlos Frenk, Volker Springel, Simon White and
the referee for helpful comments and suggestions. 
The numerical simulations analysed here were performed using the Edinburgh Cray
T3E supercomputer as part of the VIRGO consortium and partially on the 
SGI Origin at NCSA. Some of the analyses were carried out at the SGI 
Origin of the UK's National Cosmology Supercomputer COSMOS.

\end{document}